\documentclass[aps,prx,
reprint, 
amsmath,amssymb,floatfix,nofootinbib,longbibliography]{revtex4-2}

\usepackage{graphicx}
\usepackage{hyperref}
\usepackage{braket}
\usepackage{bm}
\usepackage{booktabs}
\usepackage{xcolor}
\usepackage{url}

\newcommand{\elambda}{\boldsymbol{\lambda}}

\newcommand{\half}{\tfrac{1}{2}}
\newcommand{\bbra}[1]{\langle #1|}
\newcommand{\kket}[1]{|#1\rangle}
\newcommand{\bks}[3]{\langle #1#2\Vert#3\rangle}
\newcommand{\bk}[3]{\langle #1#2|#3\rangle}

\begin{document}


\title{GW and Bethe--Salpeter Theory for Molecular Polaritons, Quasiparticles, and Excitons}

\author{Soohaeng Yoo Willow}%
\affiliation{Department of Energy Science, Sungkyunkwan University, Seobu-ro 2066, Suwon, 16419, Korea}

\author{Gi Beom Sim}%
\affiliation{Department of Energy Science, Sungkyunkwan University, Seobu-ro 2066, Suwon, 16419, Korea}

\author{Tae Hyeon Park}%
\affiliation{Department of Energy Science, Sungkyunkwan University, Seobu-ro 2066, Suwon, 16419, Korea}

\author{Tae In Kim}%
\affiliation{Department of Energy Science, Sungkyunkwan University, Seobu-ro 2066, Suwon, 16419, Korea}

\author{D. ChangMo Yang}
\affiliation{Department of Energy Science, Sungkyunkwan University, Seobu-ro 2066, Suwon, 16419, Korea}

\author{Mikul\'a\v{s} Matou\v{s}ek}
\affiliation{J. Heyrovsk\'y Institute of Physical Chemistry, Academy of Sciences of the Czech Republic, v.v.i., Dolej\v{s}kova 3, 18223 Prague 8, Czech Republic}

\author{Ji\v{r}\'i Brabec}
\affiliation{J. Heyrovsk\'y Institute of Physical Chemistry, Academy of Sciences of the Czech Republic, v.v.i., Dolej\v{s}kova 3, 18223 Prague 8, Czech Republic}

\author{Libor Veis}
\affiliation{J. Heyrovsk\'y Institute of Physical Chemistry, Academy of Sciences of the Czech Republic, v.v.i., Dolej\v{s}kova 3, 18223 Prague 8, Czech Republic}

\author{Chang Woo Myung}
\email{cwmyung@skku.edu}
\affiliation{Department of Energy Science, Sungkyunkwan University, Seobu-ro 2066, Suwon, 16419, Korea}
\affiliation{Department of Energy, Sungkyunkwan University, Seobu-ro 2066, Suwon, 16419, Korea}
\affiliation{Department of Quantum Information Engineering, Sungkyunkwan University, Seobu-ro 2066, Suwon, 16419, Korea}
\affiliation{SKKU National Lab for Intelligent Energy Solution Technology (SIEST), Sungkyunkwan University, Seobu-ro 2066, Suwon, 16419, Korea}

\date{\today}

\begin{abstract}
{The electron self-energy is the central object of quasiparticle theory, yet how an optical cavity enters it has remained an open question. 
We address this question for a molecule in a single-mode cavity using the dipole-gauge Pauli--Fierz Hamiltonian and a coherent-state QED Hartree--Fock reference. 
The cavity enters through three channels: the static dipole self-energy (DSE) contribution to
the reference orbital energies, the direct DSE augmentation of the screened interaction, and the polariton pole carrying the bilinear electron--photon coupling. 
We benchmark QED-$GW$ ionization potentials (IPs) and electron affinities (EAs)
of four ten-electron hydrides and two aromatic molecules against a cavity $\Delta$-method ladder comprising $\Delta$QED-HF, CCSD, DMRG, and polaritonic FCI. 
Where direct comparisons are possible, the correlated rungs agree within $1$~meV for the cavity-induced shifts.
The remaining $GW$ error is consistent with a missing
correlated electron--photon contribution in the charge channel whose
two states differ most in polarizability, although the energy
comparison alone does not identify a unique omitted diagram. 
For the closed-shell molecules with unbound anions studied here, the larger
error occurs in ionization: the cavity-induced redshift in IPs is
systematically overestimated. 
The cavity-induced shifts in EAs are reproduced nearly quantitatively,
but this agreement does not imply comparable accuracy for the absolute EAs or for the chemical energetics derived from them.
For strongly ionic molecules with bound anions, the ordering
reverses, consistent with published QED coupled-cluster results for sodium halides. 
Coupling and detuning scans show that the error is
predominantly quadratic in $\lambda$ and DSE-driven rather than by resonance. 
Two further consequences follow. First, the calculated spectral
function develops an isolated photoemission sideband: 
a polariton replica whose weight scales as $\lambda^2$. 
Second, in the static screened interaction used in a Bethe--Salpeter equation, 
bare-photon exchange cancels the matching DSE contribution to the direct interaction $W$. 
The complete electron--hole kernel nevertheless retains
exchange and polariton-screening corrections after this leading DSE--photon cancellation. 
Within the molecular set studied, the net effect of these corrections on the lowest excitation is appreciable only for ammonia. Because the molecular anions entering this analysis are unbound, the resulting finite-basis estimates of the exciton-binding energy are strongly basis-dependent
and should be interpreted as diagnostics of electron--hole interaction rather than as basis-converged molecular exciton-binding energies.
}
\end{abstract}

\maketitle

\section{Introduction}\label{sec:intro}

\begin{figure*}[t]
\centering
\includegraphics[width=\textwidth]{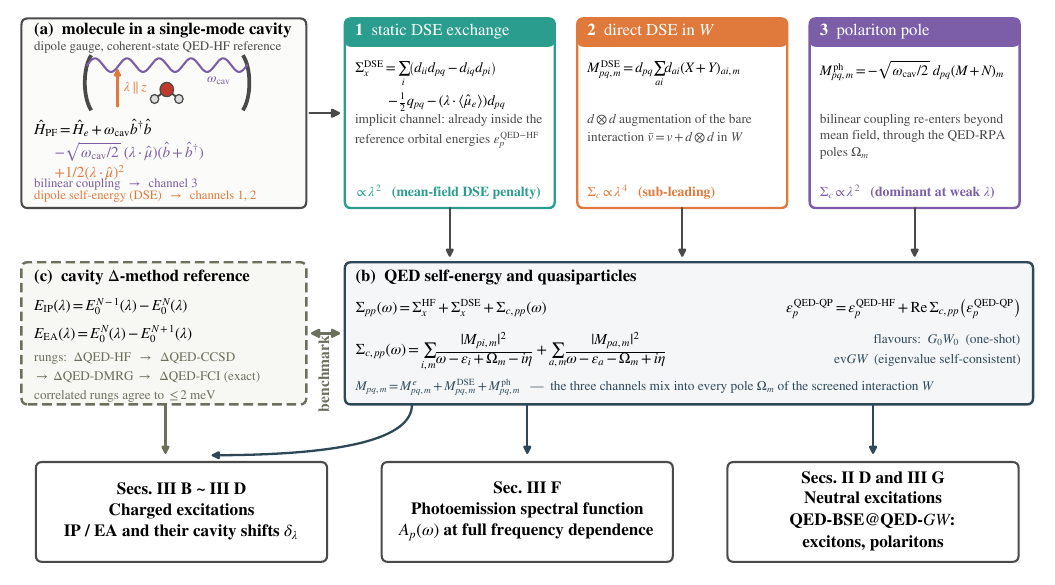}
\caption{Organization of the QED contributions to the electron
self-energy. (a) A molecule in a single-mode cavity  [Eq.~\eqref{eq:HPF}] on a
coherent-state QED-HF reference, with the coupling vector $\elambda$ polarized
along the molecular symmetry axis $z$. Its two light--matter terms
reach the self-energy through three channels. Channel~1, the static
dipole-self-energy (DSE) exchange, is \emph{implicit} in the reference orbital energies
$\varepsilon_p^\mathrm{QED\text{-}HF}$. Channels~2 and~3 are
\emph{explicit} in the transition densities [Eq.\eqref{eq:channels}]: the
direct $d\otimes d$ augmentation $M^\mathrm{DSE}$ 
and the polariton pole $M^\mathrm{ph}$. 
(b) The channels mix into every
QED-RPA pole $\Omega_m$ and assemble the
self-energy $\Sigma_c$, solved in one-shot (G$_0$W$_0$) and
eigenvalue-self-consistent (evGW) flavours. (c) The benchmark: the same charged excitations as
ground-state energy differences of the \emph{same} Pauli--Fierz
Hamiltonian on the rungs
$\Delta$QED-HF $\to$ $\Delta$QED-CCSD $\to$ $\Delta$QED-DMRG
(near-exact) $\to$ $\Delta$QED-FCI (exact).
}
\label{fig:overview}
\end{figure*}

The possibility that confining a molecule in an optical microcavity
[Fig.~\ref{fig:overview}(a)] modifies its electronic
structure even without external driving was demonstrated
experimentally by cavity-modified chemical reactivity and excited-state
dynamics under vibrational and electronic strong
coupling~\cite{Hutchison2012,Ebbesen2016,Thomas2019,GarciaVidal2021}.
At weak light--matter
coupling the cavity perturbatively dresses the electrons. At strong
coupling, polaritonic ground states emerge, and their description
requires explicit electron--photon entanglement beyond the mean-field
level. A hierarchy of \emph{ab initio} cavity quantum electrodynamics
(QED) methods now describes these polaritonic ground and excited
states~\cite{Foley2023,Ruggenthaler2023,Haugland2020,Vu2024,Yang2021,DePrince2026rpa, Bauman2025}.
Complementary reviews survey molecular-polariton dynamics and the
macroscopic-QED description of molecules in structured photonic
environments~\cite{Li2022MolecularPolaritonics,Hsu2025MacroscopicQED}.
Most of these methods build on the dipole-gauge Pauli--Fierz Hamiltonian and the
coherent-state (CS) transformation of Haugland
\emph{et al.}~\cite{Haugland2020}. They include QED-CASSCF and
QED-CASCI~\cite{Vu2024,Alessandro2025, Vu2025_casscf}, QED coupled-cluster
theory~\cite{Haugland2020,DePrince2021,Liebenthal2022,Pavosevic2022,pathak2024},
QED density-functional and TDDFT
methods~\cite{Tokatly2013,Ruggenthaler2014,Flick2017,Yang2021,Yang2022,Lu2024,Liebenthal2023}, the
QED random-phase approximation (QED-RPA)~\cite{DePrince2026rpa},
QED reduced-density-matrix methods~\cite{Mallory2022}, and QED
auxiliary-field quantum Monte
Carlo~\cite{Weber2025,Weight2024}.

Among molecular processes, \emph{charged} excitations occupy a special place.
The associated ionization potentials (IPs) and electron affinities (EAs) are
directly measurable by photoemission, and they anchor level alignment in
cavity-modified charge-transfer and redox
processes. Cavity-modulated IPs and EAs have already been predicted at
the QED coupled-cluster level~\cite{DePrince2021,Liebenthal2022}.

In electronic-structure theory the standard language for charged
excitations is the electron self-energy $\Sigma$, and its workhorse
approximation is $GW$~\cite{HedinPR1965,Strinati1980,Strinati1982,Hybertsen1985,Hybertsen1986,Godby1987,Aryasetiawan1998,Onida2002,Blase2011,vanSetten2015,Golze2019,Quintero2022,Marie2024}.
Hedin introduced the $GW$ approximation within his closed set of
equations for the one-particle Green's function, screened interaction,
polarization, self-energy, and vertex~\cite{HedinPR1965}.
The first \emph{ab initio} $GW$ calculations for a real material were
reported by Strinati, Mattausch, and Hanke for the quasiparticle band
structure of diamond~\cite{Strinati1980,Strinati1982}.
Subsequent applications to semiconductors and insulators established
$GW$ as a practical approach to quasiparticle band structures and band
gaps~\cite{Hybertsen1985,Hybertsen1986,Godby1987}; its scope has since
expanded from extended materials to molecular and nanoscale systems
and systematic molecular benchmarks~\cite{Blase2011,vanSetten2015,Golze2019}.
Extending $GW$ to cavity QED raises a conceptual question with no
counterpart in the purely electronic theory: \emph{how does the cavity
enter the self-energy?} The Pauli--Fierz Hamiltonian adds two
interactions to the electronic problem. These are the two-body dipole
self-energy (DSE) from the gauge-fixing constraint and the bilinear
electron--photon coupling. Each can appear in the reference Green's
function or in the screened interaction $W$.
Choices that look equivalent at first sight differ by entire diagram
classes. Related photon-dressed self-energies have so far been explored
mainly for extended systems, where cavity photons renormalize
electron--phonon coupling and pairing interactions~\cite{Sentef2018}; a
systematic $GW$-level organization for molecular cavity QED has been lacking.

The purpose of this paper is twofold. First, we identify and
organize the QED contributions to the self-energy at the
ring-screening level [channels~1--3 of Fig.~\ref{fig:overview}]. These
are the DSE contribution to the static
(Hartree--Fock-like) part of $\Sigma$, the direct DSE channel of $W$,
and the polariton-pole channel that carries the bilinear coupling into
the correlation self-energy $\Sigma_c$. Second, we quantify what this
ring-level self-energy captures, and what it misses. We benchmark the
resulting quasiparticle IPs and EAs of four ten-electron molecules
(H$_2$O, HF, NH$_3$, CH$_4$) and of two aromatic hydrocarbons
(benzene and naphthalene)
against a cavity $\Delta$-method [Fig.~\ref{fig:overview}(c)].
The hydrocarbons represent polarizable $\pi$-systems for which the cavity-induced corrections, and the correlated contribution missing from the ring-level self-energy, are largest;
{of the two, only benzene carries a correlated $\Delta$QED-CCSD anchor, so the quantitative statements about $\pi$-systems below are statements about benzene, with naphthalene demonstrating that the machinery reaches $68$ electrons.}
The $\Delta$-method evaluates vertical charged excitations as ground-state
energy differences of the \emph{same} Pauli--Fierz Hamiltonian for the
$N$-, $(N{-}1)$-, and $(N{+}1)$-electron systems. We take these at the
mean-field, coupled-cluster, density-matrix renormalization group (DMRG) ~\cite{White-1992, White-1993, chan_review, reiher_perspective}, and (in a minimal basis) exact
polaritonic full configuration interaction (FCI) levels. Because it
needs no diagrammatic approximation, the $\Delta$-method isolates,
channel by channel, which parts of the cavity physics the self-energy
reproduces. Finally, we extract two
observables which the self-energy delivers essentially for free, and
which no ground-state $\Delta$-method can produce. The first is the
frequency-resolved quasiparticle spectral function $A(\omega)$, whose
cavity-induced photoemission sidebands are a directly measurable prediction. 
The second is the finite-basis electron-hole gap difference conventionally written as an exciton binding energy,
$E_b = E_\mathrm{gap}-\Omega_{S_1}$, from a QED Bethe--Salpeter
equation (BSE)~\cite{SalpeterPR1951,Strinati1988,Blase2018,Blase2020,ORLANDO2023} on the QED-$GW$
quasiparticles. Because the BSE trades the QED-RPA's bare
electron--hole interaction for the attractive screened $-W$, it delivers a
finite value of this gap difference; keeping the photon explicit in the
excitation space, the polaritonic QED-BSE additionally resolves the
exciton--photon hybridization. This probes how the cavity reshapes
\emph{neutral} excitations through the same three channels.


The paper is organized as follows. Section~\ref{sec:theory} reviews
the Pauli--Fierz Hamiltonian, the coherent-state QED-HF reference, and
the photon-augmented screening. It then assembles the QED self-energy,
its quasiparticle flavours and spectral function, constructs the
QED-BSE for neutral excitations, and defines the $\Delta$-method
benchmark. Section~\ref{sec:results} presents the numerical comparison
and the decomposition of the cavity-induced IP shift. It also covers
the polariton replicas in the spectral function
(Sec.~\ref{sec:satellites}) and the cavity-modified exciton binding
(Sec.~\ref{sec:exciton}). Section~\ref{sec:conclusion} concludes.

\section{Theory}\label{sec:theory}

\subsection{Pauli--Fierz Hamiltonian and the QED-HF reference}\label{sec:pf}

We consider a molecular electronic system coupled to a single
quantized cavity photon mode of frequency $\omega_\mathrm{cav}$ in the
dipole gauge and dipole approximation. The Pauli--Fierz Hamiltonian
reads~\cite{Spohn2004,Ruggenthaler2018}
\begin{equation}
\hat{H}_\mathrm{PF}
\;=\;\hat{H}_e \;+\;\omega_\mathrm{cav}\,\hat{b}^\dagger\hat{b}
\;-\;\sqrt{\tfrac{\omega_\mathrm{cav}}{2}}\,\elambda\!\cdot\!\hat{\bm{\mu}}\,(\hat{b}+\hat{b}^\dagger)
\;+\;\tfrac{1}{2}(\elambda\!\cdot\!\hat{\bm{\mu}})^2,
\label{eq:HPF}
\end{equation}
where $\hat{H}_e$ is the bare electronic Hamiltonian, $\hat{b}$ and
$\hat{b}^\dagger$ are photon annihilation/creation operators, $\elambda$
is the cavity coupling vector (direction = polarization, magnitude =
coupling strength), and $\hat{\bm{\mu}}\,=\hat{\bm{\mu}}_e+\bm{\mu}_\mathrm{nuc}$
is the total molecular dipole operator. The last term in
Eq.~\eqref{eq:HPF} is the DSE~\cite{Haugland2020,DePrince2026rpa}. 

Following Haugland \emph{et al.}~\cite{Haugland2020}, the
coherent-state (CS) transformation~\cite{Klaudert_book} displaces the photon coordinate by
the reference dipole expectation value. This removes the linear
coupling to that expectation value at the mean-field stationary point,
while retaining the operator coupling to dipole fluctuations:
\begin{equation}
\hat{U}_\mathrm{CS}=\exp\!\Big(\tfrac{-\elambda\!\cdot\!\langle\hat{\bm{\mu}}\rangle}{\sqrt{2\omega_\mathrm{cav}}}
(\hat{b}^\dagger-\hat{b})\Big).
\label{eq:UCS}
\end{equation}
Applying this transformation to Eq.~\eqref{eq:HPF} gives
\begin{equation}
    \begin{aligned}
        \hat{H}_\mathrm{CS}
={}&\hat{U}_\mathrm{CS}\hat{H}_\mathrm{PF}\hat{U}_\mathrm{CS}^\dagger \\
={}&\hat{H}_e+\omega_\mathrm{cav}\,\hat{b}^\dagger\hat{b} 
-\sqrt{\tfrac{\omega_\mathrm{cav}}{2}}\,
\elambda\!\cdot\!\Delta\hat{\bm{\mu}}\,(\hat{b}+\hat{b}^\dagger) \\
&+\tfrac{1}{2}\bigl(\elambda\!\cdot\!\Delta\hat{\bm{\mu}}\bigr)^2,
    \end{aligned}
    \label{eq:HCS}
\end{equation}
where
$\Delta\hat{\bm{\mu}}=\hat{\bm{\mu}}-\langle\hat{\bm{\mu}}\rangle$
is the dipole-fluctuation operator and the expectation value is
evaluated with the self-consistent electronic reference. For clamped
nuclei, the nuclear dipole cancels from
$\Delta\hat{\bm{\mu}}=\hat{\bm{\mu}}_e-
\langle\hat{\bm{\mu}}_e\rangle$. 
The QED-HF reference for Eq.~\eqref{eq:HCS} is the separable
state $\ket{\Phi_0}=\ket{0_e}\otimes\ket{0_p}$, where $\ket{0_e}$ is a
Slater determinant and $\ket{0_p}$ is the photon vacuum. Equivalently,
one may work with the original Pauli--Fierz Hamiltonian and the
corresponding displaced photon state, but the Hamiltonian and state
must not both be transformed. Taking the expectation of
$\hat{H}_\mathrm{CS}$ in $\ket{\Phi_0}$ yields
\begin{equation}
E_0\;=\;\Bra{0_e}\hat{H}_e\Ket{0_e}
\;+\;\tfrac{1}{2}\Bra{0_e}\bigl(\elambda\!\cdot\![\hat{\bm{\mu}}_e-\langle\hat{\bm{\mu}}_e\rangle]\bigr)^2\Ket{0_e}.
\label{eq:EQEDHF}
\end{equation}
A crucial consequence of the CS transformation is that the bilinear electron-photon coupling has zero expectation value in the QED-HF product reference. 
This follows because both
${\bbra{0_p}(\hat{b}+\hat{b}^\dagger)\kket{0_p}=0}$ and
${\bbra{0_e}\hat{\bm{\mu}}_e-\langle\hat{\bm{\mu}}_e\rangle\kket{0_e}=0}$.
The bilinear operator itself remains present in the transformed CS Hamiltonian and couples the reference to electron-photon excited configurations; its effects therefore re-enter through the photon-pole channel of the correlation self-energy.
Thus, at the QED-HF level, the cavity modified the electronic energy only through the DSE, while dynamical electron-photon correlation appears beyond mean field.

The QED-HF Fock matrix is given by Ref.~\cite{DePrince2026rpa} as
\begin{align}
F_{pq} & = h_{pq}-(\elambda \cdot \langle\hat{\bm{\mu}}_e\rangle) d_{pq}
-\tfrac{1}{2}q_{pq} \nonumber \\
+ & \sum_i \bigl(\bks{ip}{}{iq}+d_{ii}d_{pq}-d_{iq}d_{pi}\bigr).
\label{eq:QEDHFFock}
\end{align}
The indices $i,j,\dots$ and $a, b, \dots$ denote canonical molecular spin orbitals ($\phi$) that are occupied and unoccupied, respectively, in the QED-HF reference state. 
The indices $p$ and $q$ represent general molecular spin orbitals and may refer to occupied or virtual orbitals. 
The quantity $h_{pq}$ denotes an element of the core Hamiltonian matrix.
The dipole and second moment integrals are defined as 
$d_{pq} = -\sum_\mu\lambda_\mu\Bra{\phi_p}r_\mu\Ket{\phi_q}$ and 
$q_{pq} = -\sum_{\mu\nu}\lambda_\mu\lambda_\nu \Bra{\phi_p}r_\mu r_\nu\Ket{\phi_q}$, respectively.
Finally, the antisymmetrized electron repulsion integral (ERI) is written in physicists' notation as $\bks{pq}{}{rs}\!=\!\bk{pq}{}{rs}-\bk{pq}{}{sr}$. 

Equation \eqref{eq:QEDHFFock} already contains the \emph{first} QED contribution to the self-energy.
The DSE adds two terms to the static exchange:  a
direct mean-field term $d_{ii}d_{pq}$ and an exchange term $-d_{iq}d_{pi}$.
As a result, the QED-HF orbital energies
$\varepsilon_p^\mathrm{QED\text{-}HF}$ already include the full static DSE self-energy at order $\lambda^2$,
$\Sigma_x^\mathrm{DSE}=\sum_i (d_{ii}d_{pq}-d_{iq}d_{pi})-\half q_{pq}
-(\elambda\!\cdot\!\langle\hat{\bm{\mu}}_e\rangle)d_{pq}$, on the same footing as the Fock exchange $\Sigma_x^\mathrm{HF}$. 
Because these static exchange terms already reside in $\varepsilon_p^\mathrm{QED\text{-}HF}$, 
the correlation self-energy introduced below must be constructed so as not to count them twice.


\subsection{Photon-augmented screening}\label{sec:rpa}

The correlation self-energy requires a polarization propagator~\cite{Gruneis2009,RenScheffler2012}.
The QED random-phase approximation (QED-RPA) generalizes the Casida formalism to an excitation space that includes both electronic and photonic degrees of freedom~\cite{DePrince2026rpa}.
The regrouped excitation / de-excitation basis is defined by 
$\bm{U}\!=\!(\bm{X},\bm{M})^T$, $\bm{V}\!=\!(\bm{Y},\bm{N})^T$, 
where $\bm{X}$ and $\bm{Y}$ are electronic excitation and de-excitation amplitudes and $\bm{M}$ and $\bm{N}$ are photon creation and annihilation amplitudes, respectively.
In this basis, the eigenvalue equation takes the compact form
\begin{equation}
\begin{pmatrix}\bm{\mathcal{A}} & \bm{\mathcal{B}}\\[2pt]
-\bm{\mathcal{B}} & -\bm{\mathcal{A}}\end{pmatrix}
\begin{pmatrix}\bm{U}\\[2pt]\bm{V}\end{pmatrix}
=\begin{pmatrix}\bm{U}\\[2pt]\bm{V}\end{pmatrix}\bm{\Omega},
\label{eq:QEDRPA}
\end{equation}
of dimension $2(N_\mathrm{ov}+1)$, with the photon-augmented blocks
\begin{equation}
\bm{\mathcal{A}}=\begin{pmatrix}\tilde{\bm A}&\bm g\\[2pt]\bm g^\dagger&\omega_\mathrm{cav}\end{pmatrix},\quad
\bm{\mathcal{B}}=\begin{pmatrix}\tilde{\bm B}&\tilde{\bm g}\\[2pt]\tilde{\bm g}^\dagger&0\end{pmatrix},
\label{eq:Abig}
\end{equation}
where $\tilde{\bm A} \!=\! \bm A+\bm\Delta$, $\tilde{\bm B} \!=\! \bm B+\bm\Delta'$,
$\bm{g}_{ai} \!=\! \tilde{\bm g}_{ai} \!=\! -\sqrt{\omega_\mathrm{cav}/2} ~\bm{d}_{ai}$,
and
\begin{align}
A_{ai,bj}&=(\varepsilon_a^\mathrm{QED\text{-}HF}-\varepsilon_i^\mathrm{QED\text{-}HF})\delta_{ij}\delta_{ab} \nonumber \\
& \;+\;\bk{ib}{}{aj}-\bk{ib}{}{ja},
\label{eq:A}\\
B_{ai,bj}&=\bk{ij}{}{ab}-\bk{ij}{}{ba},\label{eq:B}\\
\Delta_{ai,bj}&=d_{ai}d_{jb}-d_{ab}d_{ij},\label{eq:Del}\\
\Delta'_{ai,bj}&=d_{ai}d_{bj}-d_{aj}d_{ib}.\label{eq:Delp}
\end{align}
Here $\varepsilon_i^\mathrm{QED\text{-}HF}$ and $\varepsilon_a^\mathrm{QED\text{-}HF}$ in
Eq.~\eqref{eq:A} denote the QED-HF orbital energies of the occupied
and virtual spin orbitals. They are also the diagonal elements
${\varepsilon_p^\mathrm{QED\text{-}HF}=F_{pp}}$ of the Fock matrix of
Eq.~\eqref{eq:QEDHFFock}, and therefore already contain the static
DSE exchange of Sec.~\ref{sec:pf}.
The non-Hermitian
eigenvalue problem of Eq.~\eqref{eq:QEDRPA} has paired eigenvalues
$\pm\Omega_m$, and the positive-energy eigenvectors satisfy the
symplectic normalization
\begin{equation}
\bm{U}^T\bm{U}-\bm{V}^T\bm{V}=\bm{I}_{N_\mathrm{ov}+1}.
\label{eq:norm}
\end{equation}
The combination $\bm{U}_m+\bm{V}_m$ partitions into an electronic
block $(\bm{X} +\bm{Y})_m$ of length $N_\mathrm{ov}$ and a single
photonic amplitude $(M+N)_m$. Both enter the self-energy matrix
elements below. The eigenmodes $\Omega_m$ are \emph{polaritonic}. Away
from resonance one mode is photon-dominated and the rest
electron-dominated. But when $\omega_\mathrm{cav}$ is resonant with a
dipole-allowed excitation, as in our benchmark, the mixing is strong.

Contracted
with the bare interaction, the QED-RPA polarization yields the
screened interaction $W$.

\subsection{The QED self-energy and its two cavity channels}\label{sec:gw}

The spin-orbital $GW$@HF framework of Marie,
Ammar, and Loos~\cite{Marie2024} is adopted and extended to the photon-augmented
QED-RPA polarization propagator. The screened interaction is given in
spectral form by
\begin{align}
&W_{pq,rs}(\omega) - \bar v_{pq,rs}\nonumber\\[-1pt]
&=\sum_{m}M_{pq,m}M_{rs,m}\Big[\tfrac{1}{\omega-\Omega_m+i\eta}-\tfrac{1}{\omega+\Omega_m-i\eta}\Big],
\label{eq:Wspec}
\end{align}
where $\bar v_{pq,rs}=(pq|rs)+d_{pq}d_{rs}$ is the bare QED interaction
(Coulomb plus the dipole self-energy $d\otimes d$, chemist notation).
The transition densities are
\begin{equation}
M_{pq,m}=\!\!\!\sum_{(a,i)\!\in\!\mathrm{ov}}\!\!K_{pq,ai}\,(X+Y)_{ai,m}\;+\;K_{pq,\mathrm{ph}}\,(M+N)_m,
\label{eq:Mpqm}
\end{equation}
where the bare kernel elements that contract against the
photon-augmented QED-RPA amplitudes are
\begin{align}
K_{pq,ai}&=(pq|ai)_\mathrm{chem}+d_{pq}d_{ai},\label{eq:Kee}\\
K_{pq,\mathrm{ph}}&=-\sqrt{\omega_\mathrm{cav}/2}\;d_{pq}.\label{eq:Keph}
\end{align}
The chemist integral $(pq|ai)_\mathrm{chem}$ equals the physicist
$\bk{pa}{}{qi}$ for real spin orbitals.

Equations \eqref{eq:Mpqm}--\eqref{eq:Keph} make the channel structure
of the QED self-energy explicit. Writing the transition densities 
$M_{pq,m}=M^{e}_{pq,m}+M^\mathrm{DSE}_{pq,m}+M^\mathrm{ph}_{pq,m}$
with
\begin{align}
M^{e}_{pq,m}&=\textstyle\sum_{ai}(pq|ai)\,(X+Y)_{ai,m},\nonumber\\
M^\mathrm{DSE}_{pq,m}&=\textstyle d_{pq}\sum_{ai}d_{ai}\,(X+Y)_{ai,m},\nonumber\\
M^\mathrm{ph}_{pq,m}&=-\sqrt{\omega_\mathrm{cav}/2}\;d_{pq}\,(M+N)_m,
\label{eq:channels}
\end{align}
the cavity contributes to the dynamical self-energy through
\emph{two} explicit channels. These are the direct DSE product
$M^\mathrm{DSE}$ (the $d\otimes d$ augmentation of the screened
interaction) and the polariton-pole channel $M^\mathrm{ph}$ (the
bilinear coupling, re-entering beyond mean field). They add to the
\emph{implicit} channel already present in the reference, namely the
static DSE exchange inside $\varepsilon_p^\mathrm{HF}$
[Eq.~\eqref{eq:QEDHFFock}]. Both explicit amplitudes are proportional to
dipole matrix elements: $M^\mathrm{DSE}\propto \lambda^2$ and
$M^\mathrm{ph}\propto\lambda\sqrt{\omega_\mathrm{cav}/2}$ before the $\lambda$ dependence of the polaritonic eigenvectors is included. 
Because $\Sigma_c$ contains
$|M^e+M^\mathrm{DSE}+M^\mathrm{ph}|^2$, the interference of
$M^\mathrm{DSE}$ with $M^e$ produces a leading order-$\lambda^2$
correction ($2\mathrm{Re}[(M^e)^*M^\mathrm{DSE}] = \mathcal{O}(\lambda^2)$); only the isolated $|M^\mathrm{DSE}|^2$ term begins at
$\lambda^4$. The complete polariton-pole correction also begins at
order $\lambda^2$. Thus both cavity channels can contribute at
quadratic order, and their relative importance must be established
from the full interference-inclusive self-energy rather than from the
squares of the isolated amplitudes.
The eigenvalues $\Omega_m$ are themselves cavity-dressed through
Eq.~\eqref{eq:QEDRPA}, which mixes the two explicit channels into
every pole of $W$.

For a QED-HF reference, the Fock and DSE exchange already
reside in $\varepsilon^\mathrm{QED\text{-}HF}_p$, so the correlation
self-energy is the diagonal spectral form
\begin{align}
\Sigma_{c,pp}(\omega)&=\sum_{i\in\mathrm{occ}}\sum_m\,\frac{|M_{pi,m}|^2}{\omega-\varepsilon_i+\Omega_m-i\eta}\nonumber\\
&+\sum_{a\in\mathrm{vir}}\sum_m\,\frac{|M_{pa,m}|^2}{\omega-\varepsilon_a-\Omega_m+i\eta}.
\label{eq:Sigmac}
\end{align}
The orbital energies $\varepsilon_i$ and $\varepsilon_a$ in
the denominators of Eq.~\eqref{eq:Sigmac} are the QED-HF energies
$\varepsilon^\mathrm{QED\text{-}HF}$ at the one-shot level.
In the eigenvalue-self-consistent scheme below they are replaced by the current
quasiparticle energies, together with those entering the QED-RPA blocks of Eq.~\eqref{eq:QEDRPA}.
The quasiparticle energies $\varepsilon_p^\mathrm{QED\text{-}QP}$ are the
solutions of the diagonal quasiparticle equation
\begin{equation}
\varepsilon_p^\mathrm{QED\text{-}QP}=\varepsilon_p^\mathrm{QED\text{-}HF}
+\mathrm{Re}\,\Sigma_{c,pp}\!\left(\varepsilon_p^\mathrm{QED\text{-}QP}\right).
\label{eq:qp}
\end{equation}
Two schemes for solving Eq.~\eqref{eq:qp} are considered:
\begin{itemize}

\item \textbf{G$_0$W$_0$} (non-linear one-shot): the QED-RPA spectrum
and $M_{pq,m}$ are held fixed at the QED-HF reference, and
Eq.~\eqref{eq:qp} is solved for $\varepsilon_p^\mathrm{QED\text{-}QP}$ by Newton
iteration of the fixed point
$\omega=\varepsilon_p^\mathrm{QED\text{-}HF}+\mathrm{Re}\,\Sigma_{c,pp}(\omega)$
starting from $\omega=\varepsilon_p^\mathrm{QED\text{-}HF}$.

\item \textbf{evGW} (eigenvalue self-consistent)~\cite{Blase2011}: in each outer
iteration $n$, the photon-augmented blocks
$\tilde{\bm A},\tilde{\bm B}$ of Eq.~\eqref{eq:Abig} are rebuilt using
the current quasiparticle energies $\varepsilon^{(n)}_p$ (orbitals
themselves are fixed), the eigenvalue problem
[Eq.~\eqref{eq:QEDRPA}] is re-diagonalized, $M_{pq,m}$
[Eq.~\eqref{eq:Mpqm}] is recomputed, and Eq.~\eqref{eq:qp} is
Newton-solved for all orbitals to obtain $\varepsilon^{(n+1)}_p$ 
{(the denominators of Eq.~\eqref{eq:Sigmac} are likewise
evaluated at $\varepsilon^{(n)}_p$, while the anchor
$\varepsilon_p^\mathrm{QED\text{-}HF}$ of Eq.~\eqref{eq:qp} is kept
fixed)}, until
$\max_p|\varepsilon^{(n+1)}_p-\varepsilon^{(n)}_p|<\tau$.
\end{itemize}
The first iteration of evGW with no orbital-energy update recovers
G$_0$W$_0$. The associated quasiparticle weight is
$Z_p=\big[1-\partial_\omega\Sigma_{c,pp}|_{\omega=\varepsilon_p^\mathrm{QED\text{-}QP}}\big]^{-1}$.

Beyond its value at the quasiparticle solution, the full frequency
dependence of Eq.~\eqref{eq:Sigmac} defines the diagonal spectral
function
\begin{equation}
A_p(\omega)\;=\;-\frac{1}{\pi}\,\mathrm{Im}\,
\frac{1}{\omega-\varepsilon_p^\mathrm{QED\text{-}HF}-\Sigma_{c,pp}^{R}(\omega)},
\label{eq:Aspec}
\end{equation}
where the superscript $R$ denotes the retarded self-energy;
for a causal retarded Green function this definition guarantees a
non-negative spectral density without imposing an absolute value.
Its dominant pole is the quasiparticle peak (weight $Z_p$). Its
secondary structure consists of satellites pinned to the poles of
$\Sigma_c$, at $\omega\approx\varepsilon^\mathrm{QED\text{-}HF}_i-\Omega_m$ on the ionization
side and $\varepsilon^\mathrm{QED\text{-}HF}_a+\Omega_m$ on the affinity side: the
photoelectron (photohole) leaves behind one quantum of the excitation $m$.
In a cavity the $\Omega_m$ are \emph{polaritonic}, so
$A(\omega)$ carries photoemission replicas displaced from the parent
peak by dressed-photon rather than bare-excitation energies. This is
the cavity analogue of phonon and plasmon sidebands of photoemission spectroscopy~\cite{Aryasetiawan1996,Verdi2017}, evaluated in
Sec.~\ref{sec:satellites}.

\subsection{Neutral excitations: a QED Bethe--Salpeter
equation}\label{sec:bsetheory}

\subsubsection{QED-BSE@GW and the exciton binding energy}
More broadly, the energetics, delocalization, and dynamics of molecular
excitons have a long theoretical and spectroscopic history~\cite{Jang2018DelocalizedExcitons}.
The screened interaction that dresses the quasiparticles also defines
a Bethe--Salpeter equation for the \emph{neutral} excitations of the
cavity-coupled molecule~\cite{SalpeterPR1951,Strinati1988,AlbrechtPRL1998,RohlfingPRL1998,BenedictPRL1998,Blase2018,Blase2020,ORLANDO2023}.
{These references establish the purely electronic BSE with a
statically screened $W$; the QED generalization below inherits that
structure. Electron--photon extensions of Hedin's equations, in which
the transverse cavity photons enter the $GW$/BSE hierarchy alongside
the Coulomb interaction, were formulated by Trevisanutto and
Milletar\`i~\cite{Trevisanutto2015}, and a ``QED-BSE'' that couples
first-principles BSE excitons to quantized cavity photons has been used
to describe exciton-polaritons in two-dimensional
semiconductors~\cite{Latini2019,Novko2021}; the explicit-photon
construction of Sec.~\ref{sec:bsepol-theory} is the molecular
counterpart of that approach, built here on cavity-dressed
quasiparticles and a DSE-containing kernel. Linear-response QED
density-functional theory~\cite{Flick2019,Yang2021} and the
QED-RPA~\cite{DePrince2026rpa} are the corresponding density-functional
and bare-kernel alternatives.}
It uses the static ($\omega\!\to\!0$) limit~\cite{LoosBlase2020} of the screened interaction of
Eq.~\eqref{eq:Wspec},
\begin{align}
W^\mathrm{stat}_{pq,rs} \; & \equiv\; W_{pq,rs}(\omega{=}0) \nonumber \\
\;&=\;\bar v_{pq,rs}\;-\;2\sum_m
\frac{M_{pq,m} \; M_{rs,m}}{\Omega_m},
\label{eq:Wstat}
\end{align}
where $\bar v$ is the bare QED interaction of Eq.~\eqref{eq:Wspec} and
the coupling matrix elements $M_{pq,m}$ collect the contributions from
all three cavity channels [Eq.~\eqref{eq:channels}].
{The neutral excitations are the positive-energy solutions
of the Casida-type eigenvalue problem
\begin{equation}
\begin{pmatrix}\bm{A}^\mathrm{BSE} & \bm{B}^\mathrm{BSE}\\[2pt]
-\bm{B}^\mathrm{BSE} & -\bm{A}^\mathrm{BSE}\end{pmatrix}
\begin{pmatrix}\bm{X}^\mathrm{BSE}\\[2pt]\bm{Y}^\mathrm{BSE}\end{pmatrix}
=\begin{pmatrix}\bm{X}^\mathrm{BSE}\\[2pt]\bm{Y}^\mathrm{BSE}\end{pmatrix}
\bm{\Omega}^\mathrm{BSE},
\label{eq:BSE}
\end{equation}
which has the same structure as the QED-RPA problem of
Eq.~\eqref{eq:QEDRPA} but acts on the purely electronic particle--hole
space of dimension $2N_\mathrm{ov}$ (the photon being folded into
$W^\mathrm{stat}$) with the symplectic normalization of
Eq.~\eqref{eq:norm}. The spin-orbital blocks on top of the QED-$GW$
quasiparticles are}
\begin{align}
A^\mathrm{BSE}_{ai,bj}\;=\;&
({\varepsilon^\mathrm{QED\text{-}QP}_a}-{\varepsilon^\mathrm{QED\text{-}QP}_i})\,
\delta_{ij}\delta_{ab}\nonumber\\
&\;+\;\bk{aj}{}{ib}+d_{ai}d_{bj}\;-\;W^\mathrm{stat}_{ij,ab},
\label{eq:bse}\\
{B^\mathrm{BSE}_{ai,bj}\;=\;}&
{\bk{ab}{}{ij}+d_{ai}d_{bj}\;-\;W^\mathrm{stat}_{ib,aj},}
\label{eq:Bbse}
\end{align}
where the diagonal term of $\bm A^\mathrm{BSE}$ is the non-interacting quasiparticle energy
difference. The remaining terms constitute the BSE \emph{kernel}: the
effective electron--hole interaction (formally $\delta\Sigma/\delta G$)
that couples the transitions. This kernel is the two-particle
counterpart of the one-particle self-energy. The second and third
terms of Eq.~\eqref{eq:bse} are the bare electron--hole exchange of the QED interaction
$\bar v = v + d\otimes d$ {[$\bk{aj}{}{ib}=(ai|jb)$ in chemist notation]}. The last term, $W_{ij,ab}^\mathrm{stat}$, is the
static screened interaction of Eq.~\eqref{eq:Wstat} in the $(ij),(ab)$
channel: a chemist density pairing of an occupied--occupied density with a virtual--virtual density, distinct from the occupied--virtual transition densities $M_{pq,m}$ that build the
exchange terms.
Physically, the diagonal piece $-W^\mathrm{stat}_{ii,aa}$ is the classical screened attraction between the electron density $|\psi_a|^2$ and hole density $|\psi_i|^2$. This is what pulls excitons below the QP gaps (binding energy, optical-gap red shift, bound states below the continuum).
The coupling block $\bm B^\mathrm{BSE}$ of Eq.~\eqref{eq:Bbse} carries
the same two kernel pieces in the de-excitation channel,
which are the bare exchange $\bk{ab}{}{ij}+d_{ai}d_{bj}$ and the screened direct term $-W^\mathrm{stat}_{ib,aj}$.
$\bm B^\mathrm{BSE}$ is thus the analogue of the QED-RPA block
$\tilde{\bm B}$ of Eq.~\eqref{eq:Abig} with its bare direct term
$-\bk{ij}{}{ba}$ replaced by the screened one.

This attraction is the essential difference between the QED-BSE and the
QED-RPA. In the QED-RPA $A$ block [Eq.~\eqref{eq:A}] the electron--hole
coupling is the bare antisymmetrized interaction
$\bks{ib}{}{aj}=\bk{ib}{}{aj}-\bk{ib}{}{ja}$; the BSE keeps the exchange
$\bk{ib}{}{aj}$ but replaces the bare direct electron--hole term
$-\bk{ib}{}{ja}$ (unscreened in the QED-RPA) by the \emph{screened}
attractive interaction $-W^\mathrm{stat}_{ij,ab}$ (electron--hole
attraction). It is precisely this screened electron--hole attraction
that binds the exciton below the quasiparticle gap. 
Together with the QED-$GW$ quasiparticle gap on the diagonal, it yields the finite-basis difference
$E_b = E_\mathrm{gap} - \Omega_{S_1}$ within one framework. For the unbound molecular anions studied here this quantity
is basis-dependent and should not be interpreted as a converged
molecular exciton-binding energy.
The same rule that
prevents double counting in the self-energy (Secs.~\ref{sec:pf} and
\ref{sec:gw}) fixes the kernel: each interaction enters exactly once, the
$\bar v$ exchange bare and the direct attraction screened inside
$-W^\mathrm{stat}$, so antisymmetrizing the bare kernel on top of $W$
would count the exchange twice. In the unscreened limit
$W^\mathrm{stat}\to\bar v$, Eq.~\eqref{eq:bse} collapses to Configuration Interaction Singles (CIS) on the
QED kernel, which we verified to machine precision. Spin adaptation
follows the textbook pattern~\cite{Onida2002}: singlet excitons feel
$2[(ia|jb)+d_{ia}d_{jb}]-W$, triplets $-W$ alone. The cavity enters
through three channels that mirror Sec.~\ref{sec:gw}: the
quasiparticle energies on the diagonal (the reference channel), the
DSE $d\otimes d$ terms of the kernel, and the polaritonic poles
$\Omega_m$ of $W^\mathrm{stat}$ (the photon channel). 
{The Tamm--Dancoff approximation (TDA) drops $\bm B^\mathrm{BSE}$ and
diagonalizes $\bm A^\mathrm{BSE}$ alone; all neutral-excitation
observables reported below solve the full problem of
Eq.~\eqref{eq:BSE}.}

\subsubsection{Polaritonic QED-BSE@GW}\label{sec:bsepol-theory}
The screening-folded BSE keeps the photon implicit, inside $W$, so its
excitation space is electronic and hosts excitons but no polaritons. The
complementary construction keeps the photon \emph{explicit}, augmenting
the BSE excitation space with the cavity mode exactly as the QED-RPA does
[Eq.~\eqref{eq:Abig}] but with the screened BSE blocks in place of the
RPA ones,
\begin{equation}
\bm{\mathcal A}^\mathrm{pol}=\begin{pmatrix}\bm A^\mathrm{BSE}&\bm g\\[2pt]
\bm g^\dagger&\omega_\mathrm{cav}\end{pmatrix},\qquad
\bm{\mathcal B}^\mathrm{pol}=\begin{pmatrix}\bm B^\mathrm{BSE}&\bm g\\[2pt]
\bm g^\dagger&0\end{pmatrix},
\label{eq:bsepol}
\end{equation}
where {$\bm A^\mathrm{BSE}$ and $\bm B^\mathrm{BSE}$ are the
blocks of Eqs.~\eqref{eq:bse}--\eqref{eq:Bbse}} 
and the same light--matter vertex
$\bm g_{ai}=-\sqrt{\omega_\mathrm{cav}/2}\,d_{ai}$ as in
Eq.~\eqref{eq:Abig}. The photon now appears \emph{once}, explicitly
through $(\bm g,\omega_\mathrm{cav})$, so the screened blocks
$\bm A^\mathrm{BSE},\bm B^\mathrm{BSE}$ are built from the purely
\emph{electronic} RPA screening (the photon excluded from $W$) while the
DSE $d\otimes d$, a genuine electronic interaction, is kept. Solved as
the Casida problem of Eq.~\eqref{eq:QEDRPA}, Eq.~\eqref{eq:bsepol}
returns eigenvectors that split into an electronic block and one photonic
amplitude: at resonance the bright exciton hands its oscillator strength
to a lower and an upper polariton (Sec.~\ref{sec:abs}), with a Rabi
splitting that obeys the two-level law
$\Omega_R=\sqrt{3}\,\lambda\sqrt{f}$.
This polaritonic BSE is \emph{not}
the $\omega\!\to\!0$ limit of the screening-folded kernel: its explicit
photon enters the transition/exchange channel
($g_{ai}g_{bj}\!\sim\!d_{ai}d_{bj}$), whereas the folded photon screens
the direct channel ($d_{ij}d_{ab}$), and the two are near-orthogonal.
Equations~\eqref{eq:bse} and~\eqref{eq:bsepol} are thus distinct objects,
with the former resolving how the cavity reshapes the exciton
(Sec.~\ref{sec:exciton}) and the latter how exciton and photon hybridize
into polaritons.

{Most recently, Mauri \emph{et al.}~\cite{Mauri2026} formulated an
\emph{ab initio} BSE for exciton-polaritons that carries the photonic and
the matter components in a single excitation space and obtains polariton
dispersions at the cost of a standard static BSE calculation; their
construction is the extended-system counterpart of the explicit-photon
problem we solve in Eq.~\eqref{eq:bsepol}, with the transverse photon
entering through the same electron--hole exchange channel identified
there.}

\subsection{The $\Delta$-method benchmark}\label{sec:delta}

To assess the individual self-energy channels, 
we need charged-excitation energies obtained independently of the $GW$ self-energy truncation. 
The cavity $\Delta$-method defines the vertical ionization potential (IP) and electron affinity (EA) as
\begin{align}
E_\mathrm{IP}(\lambda)&\;=\;E_0^{N-1}(\lambda)\;-\;E_0^{N}(\lambda),
\label{eq:deltaIP}\\
E_\mathrm{EA}(\lambda)&\;=\;E_0^{N}(\lambda)\;-\;E_0^{N+1}(\lambda),
\label{eq:deltaEA}
\end{align}
where each total energy is the \emph{ground-state energy of the same Pauli--Fierz
Hamiltonian}, with identical cavity frequency $\omega_\mathrm{cav}$, light--matter coupling $\elambda$, molecular
geometry, and CS/DSE conventions, evaluated for the $N$-, $(N{-}1)$-, and $(N{+}1)$-electron systems. 
Equations \eqref{eq:deltaIP} and \eqref{eq:deltaEA} are the cavity counterparts of the conventional 
$\Delta$SCF or $\Delta$CC definitions of charged excitation energies.
At the polaritonic-FCI level, they are exact within the chosen electronic basis and photon-number truncation. 

For the closed-shell molecules considered here, the anionic state is formally unbound. 
As in any finite-basis treatment, the computed $E_\mathrm{EA}$ represents the vertical attachment energy of the anionic state represented within the finite basis. 
This provides a consistent, like-for-like reference for comparison with the self-energy based quasiparticle energies. 

A remark on gauge consistency is warranted because the
$(N{-}1)$-electron system carries a net charge, making its dipole operator origin-dependent. 
Within a fixed $N$-electron sector, however, 
shifting the coordinate origin 
changes $\elambda\!\cdot\!\hat{\bm\mu}_e$ only by a constant $c$. 
The shifted Hamiltonian is related to the original one by the photon displacement $\hat b\to\hat b - c/\sqrt{2\omega_\mathrm{cav}}$, under
which the changes in the bilinear light--matter coupling and the DSE cross term cancel exactly.
Consequently, the spectrum of each charge sector (and therefore the ionization
energy difference in Eq.~\eqref{eq:deltaIP}) is invariant with respect to the
choice of origin, provided that the full DSE, including its one-body
second-moment contribution, is retained and the photon basis is sufficiently converged. 
The same reasoning implies that each charge sector
must be assigned its \emph{own} coherent-state displacement.
Ionization changes the molecular dipole moment, causing the cavity field to relax to a new displaced vacuum.  
This change of coherent displacement is incorporated
variationally within the mean-field $\Delta$-method; dynamical
electron--photon correlation still requires a correlated rung.

Four rungs of the $\Delta$-method are used.
(i) \emph{$\Delta$QED-HF}: CS-transformed mean field for all species, with a
spin-unrestricted (QED-UHF) reference for the doublet cation and anion.
This rung contains the full static DSE physics plus the photon re-displacement,
but no dynamical correlation. The static DSE physics is described by the
mean-field counterpart of the implicit reference channel of Sec.~\ref{sec:gw}.
(ii) \emph{$\Delta$QED-CCSD}: QED
coupled-cluster with single and double excitations augmented by the
photonic amplitude classes of the QED-CCSD-1
ansatz~\cite{Haugland2020,DePrince2021}, on the QED-HF (neutral) and
QED-UHF (cation) references. {(iii) \emph{$\Delta$QED-DMRG}:
the polaritonic density-matrix renormalization group~\cite{matousek2024},
in which the cavity mode is carried as an additional site of the DMRG
lattice and all molecular orbitals are correlated; converged to the
truncation-error threshold quoted in Sec.~\ref{sec:compdet}, it is
near-exact within the cc-pVDZ basis and thereby extends the exact
anchor from the minimal basis to the production basis of the
benchmark.}
(iv) \emph{$\Delta$QED-FCI}: exact
diagonalization of $\hat H_\mathrm{PF}$ in the product space of all
electronic determinants and photon-number states $\kket{n}$, converged
in the photon truncation $n^\mathrm{max}_\mathrm{ph}$; its dense
diagonalization restricts it to a minimal basis, where it serves as
the exact anchor that calibrates rungs (ii) and (iii).

\section{Numerical results}\label{sec:results}

\subsection{Computational details}\label{sec:compdet}

Four ten-electron molecules are studied in the cc-pVDZ basis: H$_2$O
[$r_{OH}=1.0$~\AA, $\angle\mathrm{HOH}=104.5^\circ$, the geometry of
Ref.~\onlinecite{DePrince2026rpa}], HF [$r=0.917$~\AA], NH$_3$
[$r=1.0124$~\AA, $\angle\mathrm{HNH}=106.67^\circ$], and CH$_4$
[$r=1.087$~\AA, tetrahedral]. In each case the geometry is recentered
so that the nuclear centre of mass sits at the coordinate origin and
the principal symmetry axis is aligned with $z$.
A fifth hydride, PH$_3$ [$r_\mathrm{PH}=1.420$~\AA,
$\angle\mathrm{HPH}=93.345^\circ$, $C_{3v}$], enters only the
exciton-binding analysis
(Tables~\ref{tab:exciton}--\ref{tab:excitonchannel}): it is the
discriminating test of the polarization criterion of
Sec.~\ref{sec:exciton-selrule} and is not part of the IP/EA benchmark.
Two aromatic hydrocarbons, benzene and naphthalene, extend the benchmark
beyond the ten-electron set: idealized planar honeycomb geometries
($r_\mathrm{CC}=1.40$~\AA, $r_\mathrm{CH}=1.09$~\AA) in the $x$--$z$
plane with the long (spine) axis along $x$, so the $z$-polarized
cavity lies along the short in-plane axis; these centrosymmetric
geometries place the nuclear centre of mass at the origin by
construction. Benzene follows the same spin-orbital protocol as the
hydrides. For naphthalene, the quasiparticle columns use the
spin-adapted singlet implementation with density-fitted integrals and
an auxiliary-basis dielectric with density-fitted QED-UHF references for the
$\Delta$QED-HF rung.
The singlet implementation reproduces the spin-orbital quasiparticle energies to $\sim\!10^{-7}$~$E_h$ on closed-shell references.
The $\Delta$QED-CCSD rung here is omitted (the spin-orbital working set exceeds
the memory of the machines used here).
The cavity frequency is fixed at
$\omega_\mathrm{cav}=11.31$~eV for all molecules
and the coupling vector is $z$-polarized, with
$\lambda\in\{0,0.05\}$.
This makes the cavity resonant with the lowest dipole-allowed singlet excitation of \emph{water} at the standard RPA/cc-pVDZ level, and off-resonant for the others.

{The coupling $\lambda=0.05$~a.u.\ used throughout
corresponds, through $\lambda=\sqrt{1/(\epsilon_0V)}$, to an effective mode
volume of $0.74$~nm$^3$. These are single-molecule strong-coupling
results for picocavity-scale mode volume, not a simulation of a
molecular ensemble in a Fabry--P\'erot cavity, where the observed splittings
arise primarily from the collective $\sqrt{N}$ enhancement of the bright state.
Whether a particular transition is in the
ultrastrong regime should be assessed from its molecule-specific
normalized coupling $g/\omega_\mathrm{cav}$ rather than from $\lambda$
or the mode volume alone.
Appendix~\ref{ap:regime} gives the conversion and discusses
which parts of the channel decomposition are expected to transfer to the
collective regime.}

The $N$-, $(N{-}1)$-, and $(N{+}1)$-electron
calculations of Eqs.~\eqref{eq:deltaIP}--\eqref{eq:deltaEA} use the
identical molecular placement. This matters because individual
ring-level correlation energies are not
translationally invariant. The exact
$\Delta$-method differences, by
contrast, are origin-invariant, per the displacement argument of
Sec.~\ref{sec:delta}. 
Vertical IPs and EAs are reported in eV with
IP$=-\varepsilon^\mathrm{QED\text{-}QP}_\mathrm{HOMO}$,
EA$=-\varepsilon^\mathrm{QED\text{-}QP}_\mathrm{LUMO}$ for the self-energy
methods (the Koopmans approximation being the same expressions with
$\varepsilon^\mathrm{QED\text{-}HF}$ in place of $\varepsilon^\mathrm{QED\text{-}QP}$) and
Eqs.~\eqref{eq:deltaIP}--\eqref{eq:deltaEA} for the $\Delta$-method.
{The sign conventions are fixed here. 
A positive IP is the
energy required to remove an electron, whereas a positive EA is the
energy released upon attachment. Since all anions considered here are
unbound, the EAs in Table~\ref{tab:ea} are negative and are more
naturally interpreted through the corresponding positive vertical
attachment cost, $-\mathrm{EA}>0$. Accordingly, a cavity-induced shift
$\delta_\lambda\mathrm{EA}<0$ signifies destabilization of the anion
and an increase in the attachment cost.
}

$\Delta$QED-FCI anchors are computed in the STO-3G basis for H$_2$O
and HF, converged in the photon-number truncation (residual
$<10^{-4}$~eV between $n^\mathrm{max}_\mathrm{ph}=4$ and $6$). The
minimal-basis FCI dimensions of NH$_3$ and CH$_4$ exceed the dense
polaritonic solver. 
The QED-DMRG~\cite{matousek2024} energies were computed using the MOLMPS program~\cite{brabec2021,molmps}. The cavity mode was
represented as an additional lattice site with the photon-number basis
truncated at $n_{\mathrm{ph}}^{\max}=10$. All molecular orbitals were correlated, localized using the Pipek--Mezey scheme, and ordered
along the DMRG lattice using the Fiedler algorithm. The bond dimension was
dynamically adapted to satisfy a truncation-error threshold of $10^{-6}$.
{The $\Delta$QED-DMRG entries of
Tables~\ref{tab:ip}--\ref{tab:shifts} are formed from the $N$-,
$(N{-}1)$-, and $(N{+}1)$-electron DMRG ground-state energies through
Eqs.~\eqref{eq:deltaIP}--\eqref{eq:deltaEA}; they are provided for the
four hydrides in cc-pVDZ, where the dense polaritonic FCI is out of
reach.}

The absorption spectra of
Sec.~\ref{sec:abs} are built from the QED-RPA
excitation energies and electronic oscillator strengths
$f_m=\tfrac{2}{3}\,\Omega_m\sum_\kappa|\sum_{ai}\mu^\kappa_{ai}
(X{+}Y)_{ai,m}|^2$, convolved with Gaussians of width
$\sigma=0.25$~eV. The spectral functions of Sec.~\ref{sec:satellites}
evaluate Eq.~\eqref{eq:Aspec} at the one-shot (G$_0$W$_0$) level on a
frequency grid with Lorentzian broadening $\eta=5\times10^{-3}$~$E_h$.
The BSE calculations of Sec.~\ref{sec:exciton} 
solve the
\emph{full} BSE of Eq.~\eqref{eq:BSE} [blocks of
Eqs.~\eqref{eq:bse}--\eqref{eq:Bbse}] with evGW quasiparticle
energies for all orbitals, and the screening re-diagonalized at those
energies. The roots are classified as singlet or triplet by the singlet
projection $w_S=\tfrac12\sum_{AI}(T_{A\alpha I\alpha}+T_{A\beta
I\beta})^2$ of the (unit-normalized) transition amplitude $T=X+Y$
(numerically 0 or 1 throughout), and oscillator strengths are computed
from $X+Y$ in the length gauge. Section~\ref{sec:exciton}
additionally employs a second cavity tuning,
$\omega_\mathrm{cav}=7.937$~eV, resonant with the
BSE optical gap of water itself. (At $\lambda=0$ the photon
decouples, so this retuning requires no self-consistency cycle.) This
tests the detuning sensitivity of the exciton observables.

\begin{table*}[t]
\centering
\caption{Vertical ionization potential (eV) at
$\omega_\mathrm{cav}=0.415668$~$E_h$, $z$-polarized coupling
$\lambda$. Self-energy methods: IP
$=-\varepsilon^\mathrm{QED\text{-}QP}_\mathrm{HOMO}$ (Koopmans:
$-\varepsilon^\mathrm{QED\text{-}HF}_\mathrm{HOMO}$); $\Delta$-method:
Eq.~\eqref{eq:deltaIP} with doublet (QED-UHF) cations. The
$\Delta$QED-FCI anchor (dense polaritonic FCI, converged in
$n^\mathrm{max}_\mathrm{ph}$) is available in the minimal basis.
{The $\Delta$QED-DMRG column (polaritonic DMRG, all
orbitals correlated, $n^\mathrm{max}_\mathrm{ph}=10$;
Sec.~\ref{sec:compdet}) provides the near-exact reference in the
cc-pVDZ basis for the four hydrides.}
}\label{tab:ip}
\begin{ruledtabular}
\begin{tabular}{llccccccc}
system & $\lambda$ & Koopmans & $\Delta$QED-HF &
G$_0$W$_0$ & evGW & $\Delta$QED-CCSD & $\Delta$QED-DMRG &
$\Delta$QED-FCI \\
\hline
H$_2$O/STO-3G  & 0.00 & 10.584 &  8.114 &  8.853 &  8.763 &  8.457 & --- & 8.455 \\
H$_2$O/STO-3G  & 0.05 & 10.563 &  8.092 &  8.832 &  8.742 &  8.440 & --- & 8.439 \\
HF/STO-3G      & 0.00 & 12.631 & 10.710 & 11.331 & 11.293 & 11.001 & --- & 11.001 \\
HF/STO-3G      & 0.05 & 12.608 & 10.686 & 11.307 & 11.270 & 10.986 & --- & 10.986 \\
H$_2$O/cc-pVDZ & 0.00 & 13.344 & 10.562 & 12.018 & 11.911 & 11.758 & 11.791 & --- \\
H$_2$O/cc-pVDZ & 0.05 & 13.308 & 10.528 & 11.987 & 11.880 & 11.736 & 11.769 & --- \\
HF/cc-pVDZ     & 0.00 & 17.113 & 14.161 & 15.532 & 15.423 & 15.429 &  15.446 & ---\\
HF/cc-pVDZ     & 0.05 & 17.082 & 14.133 & 15.522 & 15.396 & 15.411 &  15.428 & ---\\
NH$_3$/cc-pVDZ & 0.00 & 11.402 &  9.199 & 10.585 & 10.522 & 10.287 & 10.324  & --- \\
NH$_3$/cc-pVDZ & 0.05 & 11.376 &  9.160 & 10.552 & 10.487 & 10.256 & 10.293 & --- \\
CH$_4$/cc-pVDZ & 0.00 & 14.784 & 13.326 & 14.434 & 14.394 & 14.185 & 14.2035 & --- \\
CH$_4$/cc-pVDZ & 0.05 & 14.744 & 13.227 & 14.383 & 14.353 & 14.159 & 14.1774 & --- \\
C$_6$H$_6$/cc-pVDZ & 0.00 & 9.056 & 7.847 & 9.094 & 9.077 & 9.070 & --- & --- \\
C$_6$H$_6$/cc-pVDZ & 0.05 & 8.944 & 7.730 & 9.040 & 9.019 & 9.036 & --- & --- \\
Naphthalene/cc-pVDZ & 0.00 & 7.552 & 6.629 & 7.712 & 7.702 & --- & --- & --- \\
Naphthalene/cc-pVDZ & 0.05 & 7.547 & 6.574 & 7.669 & 7.653 & --- & --- & --- \\
\end{tabular}
\end{ruledtabular}
\end{table*}

\begin{table*}[t]
\centering
\caption{Vertical electron affinity (eV), as in Table~\ref{tab:ip}
but with EA $=-\varepsilon^\mathrm{QED\text{-}QP}_\mathrm{LUMO}$ for the
self-energy methods and Eq.~\eqref{eq:deltaEA} with doublet (QED-UHF)
anions for the $\Delta$-method. All anions are unbound; 
the tabulated values are finite-basis EAs, and their negatives are the corresponding vertical attachment cose for the represented LUMO state
(the like-for-like target of the $GW$ EAs).
}\label{tab:ea}
\begin{ruledtabular}
\begin{tabular}{llccccccc}
system & $\lambda$ & Koopmans & $\Delta$QED-HF &
G$_0$W$_0$ & evGW & $\Delta$QED-CCSD & $\Delta$QED-DMRG &
$\Delta$QED-FCI \\
\hline
H$_2$O/STO-3G  & 0.00 & $-15.351$ & $-14.812$ & $-15.478$ & $-15.484$ & $-15.317$ & --- & $-15.318$ \\
H$_2$O/STO-3G  & 0.05 & $-15.390$ & $-14.826$ & $-15.497$ & $-15.503$ & $-15.336$ & --- & $-15.337$ \\
HF/STO-3G      & 0.00 & $-17.117$ & $-17.025$ & $-17.691$ & $-17.695$ & $-17.728$ & --- & $-17.728$ \\
HF/STO-3G      & 0.05 & $-17.161$ & $-17.021$ & $-17.696$ & $-17.703$ & $-17.739$ & --- & $-17.739$ \\
H$_2$O/cc-pVDZ & 0.00 & $-4.844$ & $-4.601$ & $-4.483$ & $-4.472$ & $-4.341$ & $-4.319$ & --- \\
H$_2$O/cc-pVDZ & 0.05 & $-4.980$ & $-4.681$ & $-4.560$ & $-4.547$ & $-4.415$ & $-4.394$ & --- \\
HF/cc-pVDZ     & 0.00 & $-5.003$ & $-4.775$ & $-4.774$ & $-4.768$ & $-4.722$ & $-4.728$ & --- \\
HF/cc-pVDZ     & 0.05 & $-5.200$ & $-4.836$ & $-4.838$ & $-4.830$ & $-4.784$ & $-4.790$ & --- \\
NH$_3$/cc-pVDZ & 0.00 & $-5.092$ & $-4.899$ & $-4.679$ & $-4.662$ & $-4.513$ & $-4.4728$ & --- \\
NH$_3$/cc-pVDZ & 0.05 & $-5.211$ & $-4.992$ & $-4.765$ & $-4.748$ & $-4.597$ & $-4.5563$ & --- \\
CH$_4$/cc-pVDZ & 0.00 & $-5.263$ & $-5.108$ & $-4.819$ & $-4.799$ & $-4.662$ &  $-4.626$ & --- \\
CH$_4$/cc-pVDZ & 0.05 & $-5.439$ & $-5.275$ & $-4.900$ & $-4.874$ & $-4.736$ & $-4.7009$ & --- \\
C$_6$H$_6$/cc-pVDZ & 0.00 & $-3.699$ & $-2.746$ & $-2.524$ & $-2.393$ & $-2.370$ & --- & --- \\
C$_6$H$_6$/cc-pVDZ & 0.05 & $-3.715$ & $-2.750$ & $-2.573$ & $-2.429$ & $-2.404$ & --- & --- \\
Naphthalene/cc-pVDZ & 0.00 & $-2.282$ & $-1.443$ & $-1.112$ & $-0.965$ & --- & --- & --- \\
Naphthalene/cc-pVDZ & 0.05 & $-2.390$ & $-1.508$ & $-1.181$ & $-1.013$ & --- &  --- & --- \\
\end{tabular}
\end{ruledtabular}
\end{table*}

\subsection{Quasiparticle energies versus the $\Delta$-method}\label{sec:gw-qp}

Tables~\ref{tab:ip} and \ref{tab:ea} summarize the vertical IPs and EAs
of the four hydrides and the two aromatics obtained from the Koopmans
approximation, the two
QED-$GW$ variants, and the four $\Delta$-method rungs introduced in
Sec.~\ref{sec:delta}.
For the minimal-basis calculations, the exact
$\Delta$QED-FCI results are available as benchmarks. 
In this limit, $\Delta$QED-CCSD reproduces the FCI reference to
within 2~meV for H$_2$O and below 1 meV for HF, for
both IPs and EAs at both light--matter couplings ($\lambda = 0.00$ and $0.05$). 
These results establish $\Delta$QED-CCSD as a reliable benchmark for the
systems where minimal-basis QED-FCI is computationally feasible.
{In the cc-pVDZ basis, where the dense polaritonic FCI is out
of reach, the $\Delta$QED-DMRG rung takes over the role of the
near-exact anchor for the four hydrides. Its absolute IPs lie
$17$--$37$~meV above the $\Delta$QED-CCSD values (H$_2$O $+33$, HF
$+17$, NH$_3$ $+37$, CH$_4$ $+19$~meV) and its EAs $6$--$40$~meV away
from them (H$_2$O $+22$, HF $-6$, NH$_3$ $+40$, CH$_4$ $+36$~meV).
These offsets are the residual correlations beyond singles and doubles, largely in the open-shell charge sectors.
They are the same at $\lambda=0$ and $\lambda=0.05$ to within $1$~meV, so they cancel in the cavity-induced shifts of Table~\ref{tab:shifts}; this cancellation is the decisive point for the benchmark below.}

At $\lambda=0$, the results reduce to the familiar $GW$@HF limit.
For H$_2$O, the G$_0$W$_0$ IP of $12.02$~eV reproduces the established
$GW$@HF/cc-pVDZ benchmark~\cite{vanSetten2015,Marie2024},
with a quasiparticle renormalization factor of $Z_\mathrm{HOMO}=0.946$.
Relative to the $\Delta$QED-CCSD benchmark,
evGW yields the most accurate absolute IPs 
with deviations of $+0.15$, $-0.01$, $+0.24$, $+0.21$~eV for H$_2$O,
HF, NH$_3$, and CH$_4$, respectively. 
The one-shot G$_0$W$_0$ overestimates the IPs by approximately
$0.10$--$0.30$~eV, while all methods reproduce the EAs of the unbound anions, whose values are strongly basis-dependent, within $0.05$--$0.15$~eV. {Measured
against $\Delta$QED-DMRG instead, the evGW IP deviations become
$+0.12$, $-0.02$, $+0.20$, and $+0.19$~eV and the EA deviations
$-0.15$, $-0.04$, $-0.19$, and $-0.17$~eV for H$_2$O, HF, NH$_3$, and
CH$_4$; the ranking of the methods is unchanged.}

The aromatic rows sharpen this picture for $\pi$ ionization.
For benzene, evGW deviates from $\Delta$QED-CCSD by only $+0.007$~eV
($9.077$ vs $9.070$~eV) and G$_0$W$_0$ by $+0.024$~eV
and the Koopmans value itself is nearly quantitative ($9.056$~eV, $-0.014$~eV
error), the familiar near-cancellation of orbital relaxation and correlation for
$\pi$ holes.
$\Delta$QED-HF, which contains the relaxation but not
the correlation, breaks this cancellation and underestimates the
benzene IP by $1.22$~eV.
The benzene EA shows the same compression of the $GW$ error:
evGW agrees with $\Delta$QED-CCSD to within $0.023$~eV, compared with a $0.38$~eV deviation at the $\Delta$QED-HF level.
For naphthalene, where no CCSD anchor is available, the two $GW$
flavours agree to $0.010$~eV for the IP; for the more diffuse anion they differ
by $0.15$~eV, with evGW expected to be the more reliable of the two by
extrapolation from benzene.

\begin{table*}[t]
\centering
\caption{Cavity-induced shifts
$\delta_\lambda X \equiv X(\lambda{=}0.05)-X(\lambda{=}0)$ (eV) of the
vertical IP and EA, from the
energies of Tables~\ref{tab:ip} and~\ref{tab:ea}
($\omega_\mathrm{cav}=0.415668$~$E_h$, $z$-polarized). 
These shifts isolates the cavity-induced response from the much larger absolute charged-excitation energies.
The exact $\Delta$QED-FCI
anchor is available in STO-3G and the near-exact
$\Delta$QED-DMRG anchor in cc-pVDZ, the latter reproducing $\Delta$QED-CCSD
to within $1$~meV for both observables. 
The Koopmans column is the bare
$\varepsilon^\mathrm{QED\text{-}HF}$ shift.
{Shifts are formed from unrounded energies and can differ by $1$~meV from
differences of the three-decimal entries of
Tables~\ref{tab:ip}--\ref{tab:ea}.}}\label{tab:shifts}
\begin{ruledtabular}
\begin{tabular}{lccccccc}
system & Koopmans & $\Delta$QED-HF & G$_0$W$_0$ & evGW &
$\Delta$QED-CCSD & $\Delta$QED-DMRG & $\Delta$QED-FCI \\
\hline
\multicolumn{8}{c}{\emph{Ionization potential}, $\delta_\lambda\mathrm{IP}$}\\
H$_2$O/STO-3G  & $-0.021$ & $-0.022$ & $-0.021$ & $-0.021$ & $-0.016$ & --- & $-0.016$ \\
HF/STO-3G      & $-0.022$ & $-0.025$ & $-0.025$ & $-0.022$ & $-0.014$ & --- & $-0.014$ \\
H$_2$O/cc-pVDZ & $-0.036$ & $-0.034$ & $-0.031$ & $-0.032$ & $-0.022$ & $-0.022$ & --- \\
HF/cc-pVDZ     & $-0.031$ & $-0.029$ & $-0.009$ & $-0.027$ & $-0.018$ & $-0.018$ & --- \\
NH$_3$/cc-pVDZ & $-0.026$ & $-0.039$ & $-0.034$ & $-0.035$ & $-0.031$ & $-0.031$ & --- \\
CH$_4$/cc-pVDZ & $-0.040$ & $-0.099$ & $-0.051$ & $-0.041$ & $-0.026$ & $-0.0261$ & --- \\
C$_6$H$_6$/cc-pVDZ & $-0.112$ & $-0.117$ & $-0.054$ & $-0.058$ & $-0.034$ & --- & --- \\
Naphthalene/cc-pVDZ & $-0.005$ & $-0.055$ & $-0.043$ & $-0.049$ &  --- & --- & --- \\
\hline
\multicolumn{8}{c}{\emph{Electron affinity}, $\delta_\lambda\mathrm{EA}$}\\
H$_2$O/STO-3G  & $-0.038$ & $-0.014$ & $-0.019$ & $-0.019$ & $-0.019$ & --- & $-0.019$ \\
HF/STO-3G      & $-0.044$ & $+0.005$ & $-0.005$ & $-0.008$ & $-0.011$ & --- & $-0.011$ \\
H$_2$O/cc-pVDZ & $-0.136$ & $-0.079$ & $-0.077$ & $-0.075$ & $-0.075$ & $-0.074$ & --- \\
HF/cc-pVDZ     & $-0.197$ & $-0.060$ & $-0.064$ & $-0.061$ & $-0.063$ & $-0.062$ & --- \\
NH$_3$/cc-pVDZ & $-0.119$ & $-0.094$ & $-0.087$ & $-0.086$ & $-0.084$ & $-0.084$ & --- \\
CH$_4$/cc-pVDZ & $-0.176$ & $-0.167$ & $-0.081$ & $-0.075$ & $-0.074$ & $-0.0749$ & --- \\
C$_6$H$_6$/cc-pVDZ & $-0.016$ & $-0.004$ & $-0.049$ & $-0.036$ & $-0.034$ & --- & --- \\
Naphthalene/cc-pVDZ & $-0.108$ & $-0.065$ & $-0.069$ & $-0.048$ &  --- & --- & --- \\
\end{tabular}
\end{ruledtabular}
\end{table*}

\begin{figure}[tb]
\centering
\includegraphics[width=0.98\columnwidth]{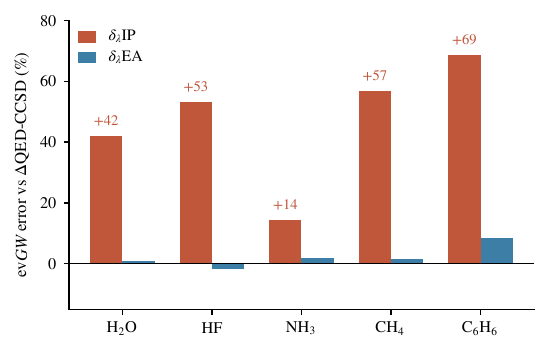}
\caption{Relative error of the evGW cavity-induced shifts against
$\Delta$QED-CCSD,
$\delta_\lambda X^{\mathrm{ev}GW}/
\delta_\lambda X^{\Delta\mathrm{QED\text{-}CCSD}}-1$, for the cc-pVDZ
entries of Table~\ref{tab:shifts} ($\lambda=0.05$); positive
means the cavity-induced shift is too large in magnitude. The two
observables behave oppositely. The ionization shift is
overestimated for every system, by $+14\%$ (NH$_3$) to $+69\%$
(benzene), whereas the attachment shift is reproduced to within $2\%$
for the hydrides and $9\%$ for benzene, and is the one channel
where the error changes sign (HF). 
Taking $\Delta$QED-DMRG as the reference instead moves the hydride bars at most one percentage point.
}
\label{fig:shifterr}
\end{figure}

\subsection{Cavity-induced shifts}\label{sec:shifts}

The \emph{cavity-induced shifts} in the polaritonic observables, $\delta_\lambda X = X(\lambda{=}0.05)-X(0)$, are summarized for all methods and molecules in  Table~\ref{tab:shifts}, and the relative error of the evGW shifts against the correlated $\Delta$QED-CCSD reference is collected in Fig.~\ref{fig:shifterr}.
For the IP of water, the resonant case, all methods predict a redshift. 
However, the $\Delta$QED-CCSD value of $-0.022$~eV (cc-pVDZ) is overestimated
by both $GW$ variants, which yield $-0.031$~eV, corresponding to an error of approximately 40\%.
The $\Delta$QED-CCSD reference is well validated: in the minimal STO-3G basis it
reproduces the exact $\Delta$QED-FCI shift ($-0.016$~eV in both methods), indicating
that the difference between the STO-3G and cc-pVDZ results arises from basis-set incompleteness rather than deficiencies in the correlation treatment. 
{The $\Delta$QED-DMRG rung closes the remaining gap in this
argument in the cc-pVDZ basis itself: its cavity-induced IP shifts
coincide with the $\Delta$QED-CCSD values to within $1$~meV for all
four hydrides ($-0.022$, $-0.018$, $-0.031$, and $-0.026$~eV for
H$_2$O, HF, NH$_3$, and CH$_4$), even though the absolute IPs of the
two rungs differ by $17$--$37$~meV. The $\sim$40\% overestimate by both
$GW$ flavours for water is therefore measured against a reference that
is near-exact in the same basis, and cannot be ascribed to the CCSD
truncation.}

A similar trend is observed across the molecular set.
The cavity-induced IP shifts $\delta_\lambda\mathrm{IP}$ are
$-0.031$~eV  versus $-0.035$~eV (evGW) for NH$_3$ ($\Delta$QED-CCSD versus evGW, $14\%$ overestimation),
$-0.026$ vs $-0.041$~eV for CH$_4$ ($+57\%$), and $-0.018$
vs $-0.027$~eV for HF ($+53\%$).
For HF the two flavours err on opposite sides of the correlated
value ($-0.018$~eV): both predict a redshift, but whereas evGW overestimates its
magnitude ($-0.027$~eV), the one-shot G$_0$W$_0$ \emph{under}estimates it,
predicting $\delta_\lambda\mathrm{IP}=-0.009$~eV, only half the
$\Delta$QED-CCSD shift.
This failure is not associated with a breakdown of the quasiparticle picture,
as the HOMO renormalization factor remains essentially unchanged
($Z_\mathrm{HOMO}=0.956$) at both light--matter couplings.
Instead, the cavity-induced shift results from the cancellation of two competing contributions: the Koopmans shift ($-0.031$~eV) and the $\lambda$
dependent change in the one-shot correlation self-energy ($\delta_\lambda \Sigma_c = +0.021$~eV).
Without self-consistency, their delicate balance is incorrectly described.
For CH$_4$, an additional complication arises from the
degenerate $t_2$ hole state of the cation.
Symmetry breaking in the QED-UHF reference produces an anomalously large mean-field shift ($-0.099$~eV), whereas $\Delta$QED-CCSD substantially restores the correct behavior
($-0.026$~eV, confirmed by $\Delta$QED-DMRG, $-0.026$~eV).

The aromatics amplify these trends. For benzene the correlated IP shift
is $\delta_\lambda\mathrm{IP}=-0.034$~eV ($\Delta$QED-CCSD), but both
mean-field estimates more than triple its magnitude (Koopmans $-0.112$, $\Delta$QED-HF $-0.117$~eV). 
The correlation correction therefore cancels approximately $70\%$ of the mean-field redshift of
the highly polarizable $\pi$-system, compared with roughly one-third for water.
The $GW$ flavours capture a substantial part of this correction, yielding shifts of $-0.054$ and $-0.058$~eV for G$_0$W$_0$ and evGW, respectively, but still
overestimate the magnitude of the correlated shift by $+58$ and $+69\%$.
This fractional error is the similar to that found for the hydrides, but occurs on a roughly three-times-larger absolute scale. 
Naphthalene shows an even stronger breakdown of the agreement between the two mean-field estimates.
The frozen-orbital Koopmans shift nearly
vanishes ($-0.005$~eV), whereas $\Delta$QED-HF gives $-0.055$~eV.
Thus, orbital relaxation and photon re-displacement of the cation, which are
negligible for water, account for essentially the entire $\Delta$QED-HF shift in naphthalene.
The two $GW$ shifts nevertheless remain close, at $-0.043$ and $-0.049$~eV for G$_0$W$_0$ and evGW, respectively. 

For benzene, the cavity-induced EA shift provides a complementary test of the mean-field picture. 
$\Delta$QED-HF substantially underestimates the correlated shift ($-0.004$ vs $-0.034$~eV for $\Delta$QED-CCSD), indicating that
the mean-field-accuracy observed for the hydride EAs does not extend to this $\pi$-system.
In contrast, evGW closely reproduces the $\Delta$QED-CCSD result, yielding
$-0.0366$ vs $-0.0338$~eV, a deviation of only $3$~meV ($+9\%$). The accuracy of the $GW$ EA shifts is
therefore not simply inherited from the mean-field cavity response; 
for the $\pi$-systems the screening contribution is essential and
is captured accurately by $GW$.\newline

\subsection{Where the self-energy succeeds and fails in cavity-induced shifts}
\label{sec:gw-bench}

The origin of the systematic overestimate becomes clear by decomposing the cavity-induced IP shift into the self-energy contributions discussed in 
Sec.~\ref{sec:gw}. The cavity correction to the IP results from the competition 
between two effects that both scale as $\lambda^2$. 
At the mean-field level, the DSE term
penalizes the state with the larger dipole fluctuations, namely the
ten-electron neutral relative to the nine-electron cation, resulting in 
a redshift of $-0.034$~eV at the $\Delta$QED-HF level. 
The corresponding Koopmans eigenvalue
shift, $-0.036$~eV, is nearly identical, indicating that cavity-induced mean-field
relaxation, including photon re-displacement in the cation, is
negligible for this system. 
Dynamical correlation partially cancels this mean-field redshift.
The Lamb-shift-like electron--photon stabilization associated with the bilinear coupling, together with the cavity-modified electron--electron
correlation mediated by the $d\otimes d$ interaction, preferentially 
stabilizes the more polarizable neutral molecule. 
These correlated contributions recover $+0.012$~eV of the
mean-field redshift, yielding the correlated $\Delta$QED-CCSD shift
of $-0.022$~eV. In contrast, the $GW$ variants recover only $+0.005$~eV through the $\lambda$ dependence of the correlation self-energy, $\Sigma_c$, 
leaving the remainder largely inherited from the implicit reference
contribution, $\varepsilon^\mathrm{QED\text{-}HF}_\mathrm{HOMO}(\lambda)$.
The ring-level self-energy thus captures only a fraction of the
correlated cavity correction present in the $\Delta$QED-CCSD energy difference. 
This discrepancy is consistent with missing vertex
and differential-correlation effects beyond ring screening. The energy
comparison alone, however, does not isolate a unique counter-term or
establish that short-range correlation is solely responsible.
Consistent with this interpretation, G$_0$W$_0$ and
evGW predict essentially identical cavity-induced IP shift for water.

\begin{figure}[tb]
\centering
\includegraphics[width=\columnwidth]{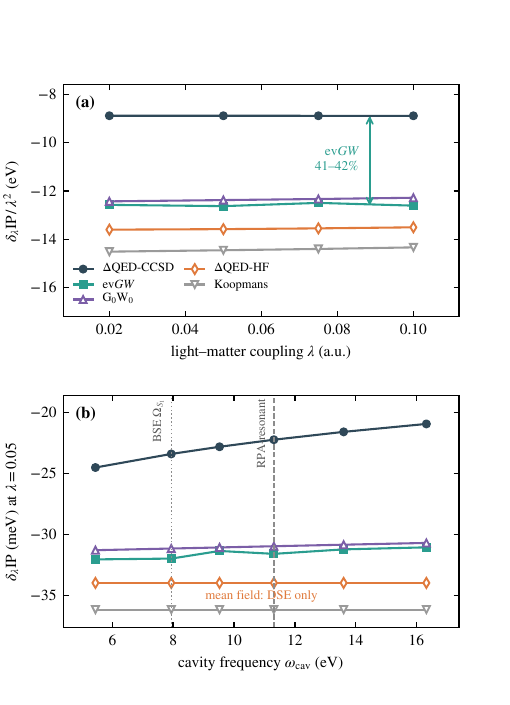}
\caption{Coupling-strength and detuning
dependence of the cavity-induced IP shift of H$_2$O/cc-pVDZ.
(a)~$\delta_\lambda\mathrm{IP}/\lambda^2$ versus $\lambda$ at
$\omega_\mathrm{cav}=11.31$~eV: a $\lambda^2$ law appears as a
horizontal line and a method's fractional error as a constant offset,
and the evGW overestimate stays at $41$--$42\%$ over a factor of five
in $\lambda$. (b)~$\delta_\lambda\mathrm{IP}$ versus
$\omega_\mathrm{cav}$ at $\lambda=0.05$: the Koopmans and
$\Delta$QED-HF shifts are exactly $\omega_\mathrm{cav}$-independent,
and the correlated shift varies monotonically with no feature at the
RPA-resonant tuning (dashed) or the BSE optical gap (dotted). The shift
is DSE-driven, not resonance-driven.}
\label{fig:ipscan}
\end{figure}

{Because both the captured and the missing
contributions scale as $\lambda^2$, their fractional difference should remain independent of light-matter coupling strength $\lambda$. Figure~\ref{fig:ipscan}(a) confirms this behavior for H$_2$O/cc-pVDZ over
$\lambda=0.02$, $0.05$, $0.075$ and $0.10$. The
$\Delta$QED-CCSD shift remains quadratic to within $1\%$, with
$\delta_\lambda\mathrm{IP}/\lambda^2=-8.9$~eV  across the entire scan.
At the same time, the evGW overestimate remains nearly constant at $+41.5$, $+42.1$, $+40.5$ and
$+41.8\%$,  even though the magnitude of the shift increases by a factor of $25$, from $-3.6$
to $-88.9$~meV. The corresponding G$_0$W$_0$ overestimate is likewise stable, varying only from $+40$ to $+38\%$.
Thus, the $\sim$40\% overestimate reported for water reflects an intrinsic feature of the
approximation rather than the particular coupling strength used for the benchmark tables.

The companion detuning scan in Fig.~\ref{fig:ipscan}(b) distinguishes
resonance-driven effects from DSE-driven physics, providing the same diagnostic used for
the BSE in Sec.~\ref{sec:exciton-selrule}. At fixed light-matter coupling of $\lambda=0.05$, we
sweep $\omega_\mathrm{cav}$ from $5.4$ to $16.3$~eV, 
spanning both the RPA resonance at $11.31$~eV and the BSE optical gap at $7.94$~eV. 
The Koopmans and $\Delta$QED-HF shifts are constant to
machine precision, at $-0.0361$ and $-0.0340$~eV, respectively, as expected
because the mean-field cavity penalty is governed by the $\omega_\mathrm{cav}$-independent
DSE operator $\tfrac12(\elambda\!\cdot\!\hat{\bm\mu})^2$ in Eq.~\eqref{eq:HPF}. 
The correlated cavity correction does depend on
$\omega_\mathrm{cav}$, but only weakly and monotonically: 
the $\Delta$QED-CCSD shift drifts from $-24.5$ to $-20.9$~meV as
$\omega_\mathrm{cav}$ increases, consistent with the $1/\omega_\mathrm{cav}$ softening of the electron--photon denominators.
Importantly, it exhibits \emph{no} feature at either resonance. 
The cavity-induced IP shift is thus predominantly an
adiabatic, DSE-driven response, even when the cavity is tuned onto a
bright electronic transition. This provides the charged-excitation counterpart to the
null result obtained for the screening-folded BSE. 
evGW is even less sensitive to detuning, varying only from $-32.0$ to $-31.0$~meV. 
It therefore captures only about one
quarter of that $\omega_\mathrm{cav}$ dependence of the correlated shift, while its fractional
error increases from $+31\%$ at $5.4$~eV to $+48\%$ at
$16.3$~eV. This behavior shows that 
evGW misses most of the observed $\omega_\mathrm{cav}$ dependence of the correlated correction, but it does not by itself identify a unique omitted diagram.
}

The same trend is
observed relative to exact $\Delta$QED-FCI in the minimal basis, which
yields cavity shifts of $-0.022$~eV (mean field), $-0.021$~eV ($GW$),
and $-0.016$~eV (FCI), thereby ruling out a basis-set artifact.
{In the cc-pVDZ basis the near-exact $\Delta$QED-DMRG shift
($-0.022$~eV) coincides with the $\Delta$QED-CCSD value, so the
$+0.012$~eV correlation correction quoted above is itself anchored to better
than $1$~meV.}

For the \emph{electron affinities}, however, the systematic overestimate found for the IP shift is essentially absent [Fig.~\ref{fig:shifterr}].
The cavity-induced EA shifts are reproduced almost quantitatively by all $GW$ variants
(evGW vs $\Delta$QED-CCSD:
$-0.075$ vs $-0.075$~eV for H$_2$O, $-0.061$ vs $-0.063$~eV for HF,
$-0.086$ vs $-0.084$~eV for NH$_3$, $-0.075$ vs $-0.074$~eV for
CH$_4$). $\Delta$QED-DMRG agrees with $\Delta$QED-CCSD to
within $1$~meV here as well ($-0.074$, $-0.062$, $-0.084$, and
$-0.075$~eV), so the agreement of the $GW$ attachment shifts holds
against the near-exact cc-pVDZ reference.
Upon electron attachment, the correlated correction is small in this basis
because the added electron occupies a diffuse orbital that is artificially bound by the finite Gaussian basis, with weak differential dipole-fluctuation correlation.
Consequently, $\Delta$QED-HF
already lies within $\sim$10\% of $\Delta$QED-CCSD, and a self-energy 
inheriting the mean-field cavity response remains accurate. 
These EA shifts are vertical and basis-dependent, 
since the anion is artificially bound by the finite basis, 
and their absolute \emph{magnitudes} therefore depend on the
basis set. For example, adding diffuse functions increases the water
$\Delta$QED-CCSD shift from $-0.075$~eV (cc-pVDZ) to $-0.235$~eV
(aug-cc-pVDZ) as the excess electron becomes more delocalized.
Crucially, however, the \emph{agreement} between $GW$ and CCSD survives this basis change
(aug-cc-pVDZ: $-0.241$ vs $-0.235$~eV, within $3\%$), demonstrating that it
reflects a common description of the cavity response rather than a shared basis artifact.
In contrast, the proximity between the mean-field and CCSD shifts is specific to the cc-pVDZ basis.
With diffuse functions, $\Delta$QED-HF predicts a substantially larger shift ($-0.363$~eV), indicating that the
differential correlation grows with the diffuseness of the attached electron. 

{
This near-quantitative agreement applies specifically to the differential response $\delta_\lambda\mathrm{EA}$. 
The baseline errors in the absolute evGW EAs ($0.05$--$0.15$~eV relative to $\Delta$QED-CCSD) are comparable to or larger than the cavity-induced shifts themselves. 
Although much of this baseline error cancels when the same molecule is compared at $\lambda=0$ and finite $\lambda$, it remains relevant to absolute attachment thresholds, level alignment, redox energetics, and other chemical predictions. 
Thus, the present benchmark establishes accuracy for cavity-induced EA changes, 
whereas predictive absolute chemical energetics additionally require control of the zero-coupling EA and of the finite-basis description of the anion.}
The central conclusion therefore remains unchanged: 
$GW$ accurately captures cavity-induced attachment shift, 
whereas its failure is specific to ionization, precisely
where the correlated $\lambda^2$ correction becomes large. 
The raw Koopmans estimate (e.g.\ $-0.136$~eV for
water) also overestimates the EA shift by a factor of two.
In this case, orbital relaxation of the anion,
captured by both $\Delta$QED-HF and the $GW$ screening, 
restores  agreement with $\Delta$QED-CCSD.

The QED-$GW$ route is proposed to estimate the screened interaction $W$, 
which is not constructed in the coupled-cluster route and is subsequently used in Secs.~\ref{sec:abs}--\ref{sec:exciton-basis} 
to describe neutral excitations, the electron--hole kernel, and spectral satellites. 
The IP/EA benchmarks in Tables~\ref{tab:ip}--\ref{tab:shifts} thus serve as a fidelity check on $W$ 
in a charged-excitation channel for which independent correlated reference data are available. 
Appendix~\ref{ap:sodium_halides} performs this check directly against the published QED-CC results, 
reproducing the setup of Ref.~\onlinecite{DePrince2021} for NaF and NaCl. 
The results are consistent with a missing correlated contribution being largest in the charge channel 
whose two states differ most strongly in polarizability,
which is attachment rather than ionization for these ionic species; 
they do not uniquely identify its diagrammatic form.

\begin{figure*}[tb]
\centering
\includegraphics[width=0.8\textwidth]{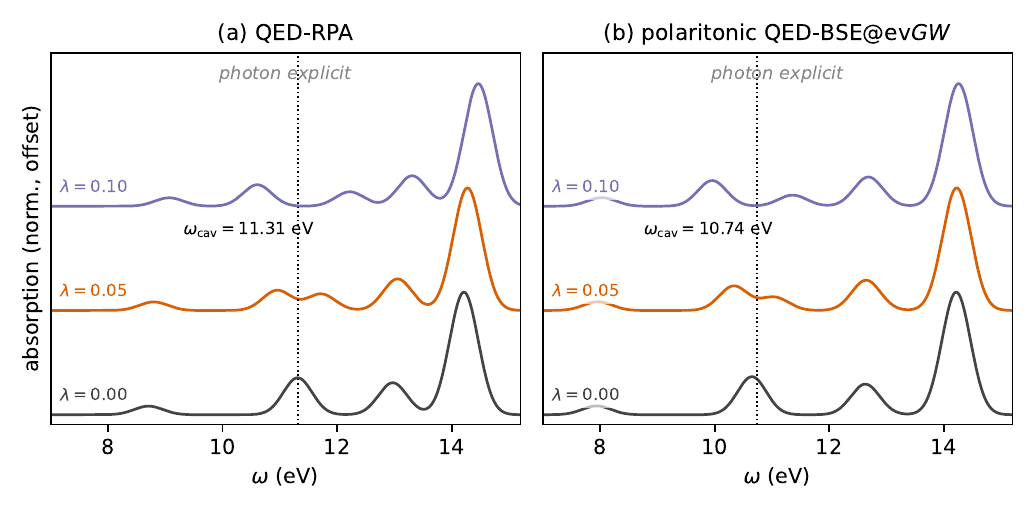}
\caption{Cavity-modified absorption of H$_2$O/cc-pVDZ at $\lambda=0$,
$0.05$, and $0.10$ (normalized to the tallest peak, offset
vertically; Gaussian broadening $0.25$~eV). 
(a)~QED-RPA, tuned to the lowest dipole-allowed RPA root
($\omega_\mathrm{cav}=11.31$~eV).
(b)~polaritonic QED-BSE@evGW [Eq.~\eqref{eq:bsepol}],  tuned to the bright
BSE singlet at $\omega_\mathrm{cav}=10.74$~eV (dotted lines). In both
panels the photon is explicit, so the cavity-resonant band Rabi-splits, by $0.77 \to 1.62$~eV in (a) and
$0.71 \to 1.41$~eV in (b).
The QED-RPA poles $\Omega_m$ of panel (a) are those entering $W$ [Eq.~\eqref{eq:Wspec}].}\label{fig:abs}
\end{figure*}

\begin{figure*}[tb]
\centering
\includegraphics[width=0.8\textwidth]{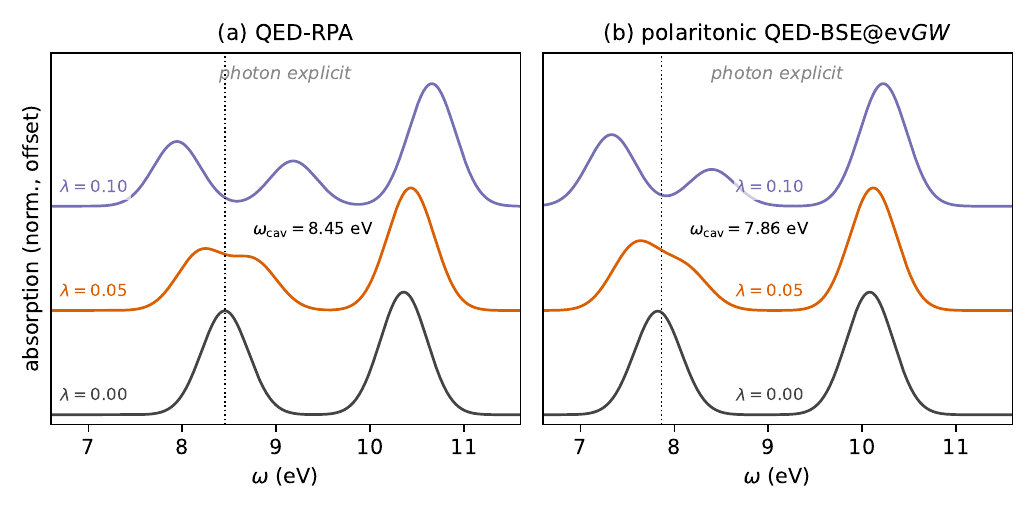}
\caption{As Fig.~\ref{fig:abs}, for NH$_3$/cc-pVDZ and zoomed to the
cavity-resonant band. In $C_{3v}$ the $z$-polarized cavity couples only
to the $z$-polarized $A_1$ transitions, so it is tuned to the lowest
dipole-allowed $A_1$ root (RPA $\omega_\mathrm{cav}=8.45$~eV; BSE
analogue $7.86$~eV, dotted lines). This $A_1$ band Rabi-splits in both
explicit-photon panels; the
neighbouring $\sim10$~eV degenerate $E$ ($x,y$) feature, dark to the
$z$-cavity, is unshifted and serves as a reference.}\label{fig:absnh3}
\end{figure*}

\begin{figure*}[tb]
\centering
\includegraphics[width=0.8\textwidth]{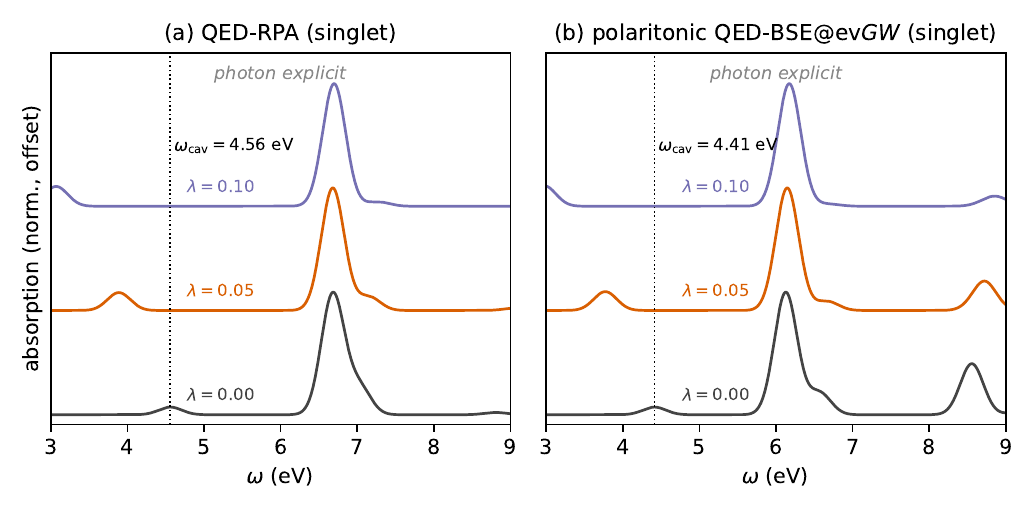}
\caption{As Fig.~\ref{fig:abs}, for naphthalene/cc-pVDZ in
the spin-adapted singlet implementation (density-fitted
integrals, auxiliary-basis dielectric for the evGW step, Gaussian broadening $0.15$~eV). 
The cavity is polarized along the short in-plane
axis ($z$) and tuned to the lowest bright singlet, corresponding to
the weak short-axis transition at $\omega_\mathrm{cav}=4.56$~eV (RPA) /
$4.41$~eV (BSE, dotted), far below the dominant long-axis band
at $6.7$/$6.1$~eV. Only the \emph{lower} polariton appears, 
red-shifting and brightening with $\lambda$, while the upper branch stays dark and the long-axis
band is essentially
unshifted.}\label{fig:absnaph}
\end{figure*}

\subsection{Polaritonic structure of the screening: absorption
spectrum}\label{sec:abs}

It is instructive to visualize the poles $\Omega_m$ that enter the
screened interaction $W$ of Eq.~\eqref{eq:Wspec}, which the quasiparticle
analysis of the preceding sections accessed only through the
correlation self-energy. Figure~\ref{fig:abs} shows the low-frequency
QED-RPA absorption spectrum of H$_2$O/cc-pVDZ at $\lambda=0.00$, $0.05$, and $0.10$. 
This provides the cavity analogue of the RPA spectra of Ref.~\onlinecite{Marie2024}. 
At $\lambda=0$ the photon mode is
optically dark and the spectrum is the bare RPA spectrum, with the
lowest dipole-allowed band at $11.31$~eV coinciding with
$\omega_\mathrm{cav}$ by construction. At finite coupling this
resonant band Rabi-splits into a lower and an upper polariton: at
$\lambda=0.05$ they appear at $10.95$ and $11.72$~eV with photon-coordinate strengths
$|(M{+}N)_m|^2=0.65$ and $0.36$, respectively, and the
splitting grows from $0.77$~eV to $1.62$~eV at $\lambda=0.10$. The
remaining bands shift only weakly (the $\lambda^2$ DSE blue-shift).

Panel (b) recasts the same spectrum at the correlated
polaritonic QED-BSE@evGW level [Eq.~\eqref{eq:bsepol}, Sec.~\ref{sec:exciton}],
with the photon kept explicit in the BSE excitation space on the
eigenvalue-self-consistent evGW quasiparticle reference; as in QED-RPA,
the bright band Rabi-splits into lower and upper polaritons.
Fig.~\ref{fig:absnh3} repeats the comparison for NH$_3$, whose
$z$-polarized cavity is resonant with its lowest $A_1$ exciton.
These polaritonic poles are exactly the structures that the
photon-pole channel $M^\mathrm{ph}$ of Eq.~\eqref{eq:channels} feeds
into $\Sigma_c$. The figure thus displays the screening input of the
QED self-energy in an observable, and makes clear that this input
is strongly \emph{frequency-structured} around $\omega_\mathrm{cav}$.

Naphthalene (Fig.~\ref{fig:absnaph}) extends the comparison to a
68-electron aromatic, computed with the spin-adapted singlet
implementation of both solvers (density-fitted integrals and an
auxiliary-basis dielectric for the evGW step; validated against the
spin-orbital modules on closed-shell references), whose excitation
space is the spatial singlet block rather than the spin-orbital manifold. 
Here the cavity is polarized along the short in-plane axis and
tuned to the lowest bright singlet, corresponding to the \emph{weak} short-axis
transition ($f\simeq0.10$ at $4.56$~eV RPA / $4.41$~eV BSE), 
rather than to a dominant band as in H$_2$O and NH$_3$.
Consequently, the polaritonic response becomes strikingly asymmetric. 
The lower polariton disperses
from $\omega_\mathrm{cav}$ down to $3.89$ ($3.07$)~eV at
$\lambda=0.05$ ($0.10$) at the RPA level [$3.77$ ($2.98$)~eV at the
BSE level] and more than doubles its oscillator strength
($f=0.10\to0.23\to0.26$), while \emph{no upper-polariton peak emerges
at all}: every root between $\omega_\mathrm{cav}$ and the first
electronic band stays dark ($f<0.002$). The two-level Rabi picture,
which distributes the resonant exciton's strength over both branches
(as observed for H$_2$O and NH$_3$, whose cavities sit on strong
transitions), fails here because the photon couples simultaneously to
\emph{all} short-axis singlets. In the upper branch the admixtures of
the $4.6$~eV exciton and the higher short-axis states enter out of
phase and their transition dipoles interfere destructively; in the
lower branch the same admixtures add constructively. The lower
polariton therefore brightens beyond the bare exciton precisely at the
expense of the next short-axis band at ${\sim}7.0$~eV, whose strength
collapses from $f=0.36$ to $0.06$ over the same $\lambda$ scan. The
dominant long-axis band at $6.7$~eV ($6.1$~eV BSE), dark to the
$z$-polarized cavity, keeps $f\simeq1.6$ throughout and serves as the
unsplit internal reference, exactly as the $E$ band of NH$_3$ in
Fig.~\ref{fig:absnh3}. Polariton-branch brightness in a molecular
cavity is thus a multi-state interference property of the electronic
structure, not a property of the resonant transition alone.

Two remarks clarify the underlying physics.
First, the screening-folded BSE of Eq.~\eqref{eq:bse} is formulated entirely within  electronic excitation space.
Its eigenvalues therefore describe
\emph{excitons} dressed by the polaritonic screening rather than the
lower and upper polaritons themselves.
The polaritons appear as the poles of
the screening (Fig.~\ref{fig:abs}) entering the BSE kernel.
By contrast, retaining the
photon as an explicit degree of freedom yields the
polaritonic QED-BSE of Eq.~\eqref{eq:bsepol},
which recovers the lower and upper polariton branches directly.
The BSE thus isolates the effect of the cavity on the electronic excitation spectrum. 
{Within the same diagrammatic framework, it provides the quasiparticle gap,
\begin{equation*}
E_\mathrm{gap}=\varepsilon^\mathrm{QED\text{-}QP}_\mathrm{LUMO}
-\varepsilon^\mathrm{QED\text{-}QP}_\mathrm{HOMO},
\end{equation*}
where the LUMO and HOMO are defined as the lowest quasiparticle energy among the virtual orbitals and the highest among the occupied orbitals, respectively. 
This distinction is important because evGW shifts individual orbitals independently. In particular, when the cavity splits the degeneracy of a frontier manifold, as for the $t_2$ levels of CH$_4$, an index-based assignment can select the wrong member of the multiplet and alter $\delta_\lambda E_\mathrm{gap}$ by several meV.
The same framework also yields the optical gap
$\Omega_{S_1}$, defined by the lowest singlet excitation, and hence the exciton binding energy, 
\begin{equation}
E_b\;=\;E_\mathrm{gap}\;-\;\Omega_{S_1}.
\label{eq:Eb}
\end{equation}
Together, these quantities provide a unified description connecting quasiparticle and optical excitations.}
This simultaneous access to quasiparticle and excitonic observables is a distinctive advantage of the $GW$+BSE framework among cavity-QED approaches.

Second, the static limit exhibits an exact but index-specific cancellation. Eliminating a linearly
coupled bare photon at $\omega=0$ generates 
$-2\,g_{pq} g_{rs}/\omega_\mathrm{cav} =
-\,d_{pq} d_{rs},$ 
which cancels the matching DSE contribution $+d_{pq} d_{rs}$ 
inside the direct screened interaction $W_{pq,rs}^\mathrm{stat}$ [Appendix \ref{ap:cancel}].
This identity is a property of the direct
interaction at the same four indices; it does not remove the separate
bare-exchange DSE term $d_{ai}d_{bj}$ in
Eqs.~\eqref{eq:bse}--\eqref{eq:Bbse}. The complete BSE kernel is
therefore not, in general, identical to the purely electronic kernel.
For the low-lying states examined below, the matching terms inside
$W$ cancel closely and the remaining exchange and polariton-screening
corrections are often small, producing the numerical
``cavity-transparency'' reported in Sec.~\ref{sec:exciton-selrule}.
Whether the residual reaches a particular exciton is controlled by the
symmetry and dipole structure of that state, rather than by the bare
static identity alone.
%

\begin{figure}[!htbp]
\centering
\includegraphics[width=0.95\columnwidth]{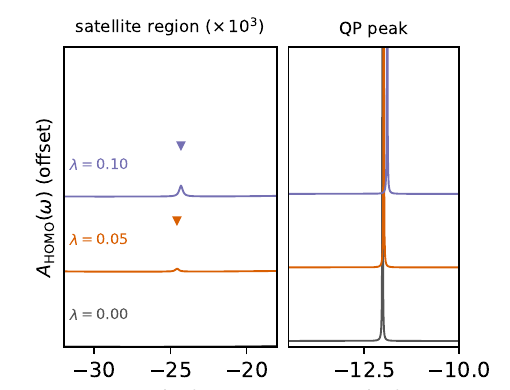}
\caption{HOMO spectral function of H$_2$O/cc-pVDZ
[Eq.~\eqref{eq:Aspec}, G$_0$W$_0$ level, $\eta=5\times10^{-3}$~$E_h$]
at $\lambda=0$, $0.05$, $0.10$ (curves offset vertically). Right:
quasiparticle peak ($Z\simeq0.94$). Left: the satellite window,
magnified $\sim\!10^3$-fold relative to the right panel. The bare
molecule ($\lambda=0$) has no spectral weight in this window; at
finite coupling the polariton replica appears at
$\varepsilon^\mathrm{QED\text{-}HF}_\mathrm{HOMO}-\tilde\omega$ (triangles mark the pole
positions of $\Sigma_c$), with weight $\propto\lambda^2$.}
\label{fig:spec}
\end{figure}

\subsection{Polariton replicas in the photoemission spectral
function}\label{sec:satellites}

The polariton-pole contribution to $\Sigma_c$, which contributes only
weakly to the cavity-induced quasiparticle shifts
(Sec.~\ref{sec:gw-bench}), becomes directly
visible once the self-energy is resolved in frequency.
Figure~\ref{fig:spec} shows the HOMO spectral function
[Eq.~\eqref{eq:Aspec}] of H$_2$O/cc-pVDZ at $\lambda=0$, $0.05$, and
$0.10$. The quasiparticle peak at $-12.0$~eV carries a spectral weight of
$Z\simeq 0.94$ and exhibits the quasiparticle redshift of $\delta_\lambda \mathrm{IP} \approx -0.031$ eV. 
In contrast, the present G$_0$W$_0$ calculation has no resolved bare-molecule intensity in the satellite window between $-18$ and $-30$~eV.
By symmetry, only excitations with totally symmetric transition densities
contribute to the HOMO diagonal element of the self-energy, 
and the corresponding intrinsic RPA shake-up satellites,
$\varepsilon^\mathrm{QED\text{-}HF}_i-\Omega_m$, occur at substantially lower energies.
Upon introducing cavity coupling, a new satellite emerges within this otherwise empty
window at $\omega=\varepsilon^\mathrm{QED\text{-}HF}_\mathrm{HOMO}-\tilde\omega$.
Both families of poles are referenced to the QED-HF eigenvalues, since the one-shot $\Sigma_c=iG_0W$ inherits the poles of the QED-HF Green function $G_0$.
The occupied branch sits at $\varepsilon^\mathrm{QED\text{-}HF}_i - \Omega_m$, of which the polariton replica is the special case $i=\mathrm{HOMO}$.
It appears at $-24.56$~eV for $\lambda=0.05$ and  shifts slightly to $-24.30$~eV for $\lambda=0.10$. 
Here, $\tilde\omega$ denotes the photon-dominated polariton of the QED-RPA screening,
with energy  $11.26$~eV at $\lambda=0.05$ and photon-coordinate strength $|(M{+}N)_m|^2\approx1.0$.
This large photon-coordinate response reflects the near-resonant tuning:
the cavity mode lies just below the lowest bright polariton of the
screening, so the pole entering the diagonal self-energy is essentially
the bare-photon replica.
This mode is distinct from the two \emph{bright} polaritons
of Fig.~\ref{fig:abs} ($10.95$ and $11.72$~eV at $\lambda=0.05$), between
which it lies: it is optically dark ($f<10^{-3}$) and therefore absent
from the absorption spectrum, yet it is the pole that dominates the
\emph{diagonal} self-energy, because the photonic contribution to
$M_{pp,m}$ [Eq.~\eqref{eq:channels}] is proportional to
$d_{pp}\,(M{+}N)_m$ and thus samples the photon amplitude of a mode
irrespective of its oscillator strength.
This satellite corresponds to the photoemission accompanied by
creation of one dressed-photon quantum in the cavity, serving as the polaritonic
counterpart of a phonon sideband. Its pole strength
increases from $|M_{pp,m}|^2=1.5\times10^{-5}$ to $5.2\times10^{-5}$
$E_h^2$ as $\lambda$ increases from $0.05$ to $0.10$.
The resulting enhancement by a factor of $3.4$, close to the factor
of $4$ expected from $\lambda^2$ scaling, confirms the quadratic
dependence on the light--matter coupling, arising from the linear
dependence $M^\mathrm{ph}\propto\lambda$.
Consistently, the integrated sideband weight (above the $\lambda=0$
broadening background of the window) grows from $0.6\times10^{-4}$ to
$2.0\times10^{-4}$.
Because the cavity frequency $\omega_\mathrm{cav}$ is resonant 
with the electronic spectrum, the sideband is displaced by 
$\tilde\omega\simeq 11$~eV from the QED-HF parent level, 
corresponding to a separation of $12.6$~eV from the renormalized quasiparticle peak in Fig.~\ref{fig:spec}.
The additional $1.6$~eV reflects the difference $\varepsilon^\mathrm{QP} - \varepsilon^\mathrm{QED\text{-}HF}$,
placing the sideband in a spectral region 
where the bare molecule has essentially no spectral intensity.
Within the present G$_0$W$_0$ calculation it therefore provides an isolated,
$\lambda^2$-calibrated photoemission
signature of the bilinear light--matter interaction that complements 
optical absorption and should, in principle, be measurable by cavity
photoelectron spectroscopy. 

The conceptual message is noteworthy.
The very \emph{same} pole of the screened interaction $W$
that contributes only weakly to the quasiparticle energies
(Sec.~\ref{sec:gw-bench}) becomes directly observable
when the full frequency dependence of the self-energy is retained.
The spectral function therefore reveals cavity physics  at
the polariton frequency that is not captured by a quasiparticle-energy analysis alone.

\begin{table}[!htbp]
\centering
\caption{Neutral-excitation observables from QED-BSE@QED-evGW@QED-HF
(full BSE, cc-pVDZ, $\omega_\mathrm{cav}=0.415668$~$E_h$, $z$-polarized
coupling): fundamental quasiparticle gap $E_\mathrm{gap}$, lowest
singlet (optical gap) $\Omega_{S_1}$ and lowest triplet
$\Omega_{T_1}$ excitation energies, singlet--triplet gap
$\Delta E_\mathrm{ST}=\Omega_{S_1}-\Omega_{T_1}$, and exciton binding
energy $E_b=E_\mathrm{gap}-\Omega_{S_1}$, in eV. The
$\delta_\lambda$ rows give the cavity-induced shifts
$X(\lambda{=}0.05)-X(0)$.}\label{tab:exciton}
\begin{ruledtabular}
\begin{tabular}{llccccc}
system & $\lambda$ & $E_\mathrm{gap}$ & $\Omega_{S_1}$ &
$\Omega_{T_1}$ & $\Delta E_\mathrm{ST}$ & $E_b$ \\
\hline
H$_2$O & 0.00 & 16.383 & 7.937 & 7.116 & 0.820 & 8.447 \\
       & 0.05 & 16.427 & 7.970 & 7.143 & 0.826 & 8.457 \\
       & $\delta_\lambda$ & $+0.044$ & $+0.033$ & $+0.027$ & $+0.006$ & $+0.011$ \\
HF     & 0.00 & 20.191 & 10.837 & 10.080 & 0.757 & 9.354 \\
       & 0.05 & 20.226 & 10.855 & 10.092 & 0.763 & 9.371 \\
       & $\delta_\lambda$ & $+0.034$ & $+0.018$ & $+0.013$ & $+0.005$ & $+0.016$ \\
NH$_3$ & 0.00 & 15.184 & 7.823 & 7.090 & 0.733 & 7.361 \\
       & 0.05 & 15.235 & 7.881 & 7.129 & 0.752 & 7.354 \\
       & $\delta_\lambda$ & $+0.051$ & $+0.058$ & $+0.039$ & $+0.019$ & $-0.007$ \\
CH$_4$ & 0.00 & 19.193 & 12.532 & 11.268 & 1.264 & 6.661 \\
       & 0.05 & 19.227 & 12.556 & 11.282 & 1.274 & 6.671 \\
       & $\delta_\lambda$ & $+0.034$ & $+0.023$ & $+0.013$ & $+0.010$ & $+0.010$ \\
PH$_3$ & 0.00 & 14.499 & 8.133 & 6.207 & 1.926 & 6.366 \\
       & 0.05 & 14.531 & 8.140 & 6.199 & 1.941 & 6.391 \\
       & $\delta_\lambda$ & $+0.032$ & $+0.007$ & $-0.008$ & $+0.015$ & $+0.025$ \\
\end{tabular}
\end{ruledtabular}
\end{table}

\begin{figure*}[tb]
\centering
\includegraphics[width=0.86\textwidth]{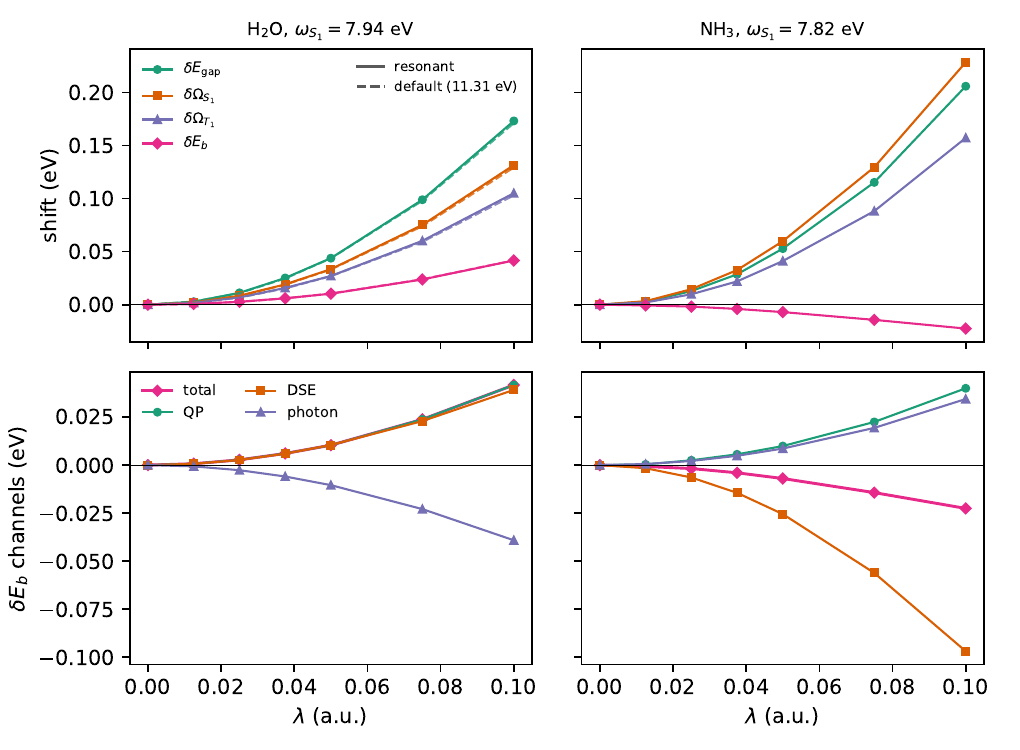}
\caption{Channel decomposition of the cavity-induced exciton-binding
shift for H$_2$O (left) and NH$_3$ (right), each with the cavity tuned
to its own BSE optical gap ($S_1$-resonant:
$\omega_\mathrm{cav}=7.94$ and $7.82$~eV; full QED-BSE@QED-evGW, cc-pVDZ, $z$-polarized). 
Top: cavity-induced shifts of 
$\delta_\lambda E_\mathrm{gap}$, $\delta_\lambda\Omega_{S_1}$,
$\delta_\lambda\Omega_{T_1}$, and 
$\delta_\lambda E_b$ versus $\lambda$; for water the default
RPA-resonant cavity ($11.31$~eV, dashed) is overlaid on the
$S_1$-resonant scan (solid) and coincides with it to $\le2\%$.
Bottom: the additive QP/DSE/photon
channels of $\delta_\lambda E_b$ (cf.\ Table~\ref{tab:excitonchannel}).
For H$_2$O the DSE and photon channels cancel; 
for NH$_3$, whose $S_1$ is bright along the
cavity axis, they do not.
Both panels are cc-pVDZ, where the residual dominates; it does not in aug-cc-pVDZ (Sec.~\ref{sec:exciton-basis}).
}
\label{fig:bindcompare}
\end{figure*}

\subsection{Cavity-modified exciton binding from the QED-BSE: observables and trends}\label{sec:exciton}

Cavity engineering of excitons, including their binding, dispersion, and
hybridization with cavity photons, has previously been explored from
first principles for two-dimensional
semiconductors~\cite{Latini2019,Novko2021}; the QED-BSE@QED-$GW$ framework
developed here extends this question to molecules, with the cavity
incorporated self-consistently through the quasiparticle energies, the DSE
kernel, and the polaritonic screening.

Table~\ref{tab:exciton} collects the neutral (or optical)-excitation observables
of five molecules from the full QED-BSE of Eq.~\eqref{eq:bse} on top of evGW
quasiparticles: the fundamental gap $E_\mathrm{gap}$, the optical gap
$\Omega_{S_1}$, the lowest triplet $\Omega_{T_1}$, the
singlet--triplet gap, and the finite-basis difference $E_b$ [Eq.~\eqref{eq:Eb}].
For these compact molecules in a confined Gaussian basis,
$E_b$ is large, $6.4$--$9.4$~eV. 
Because the anions are unbound, both the fundamental gap and this difference depend strongly on the representation of the attachment state.
The cavity-induced shifts below are therefore finite-basis electron--hole
diagnostics; their basis dependence is quantified in
Sec.~\ref{sec:exciton-basis}.
For H$_2$O, the cavity increases the fundamental gap by $+0.044$~eV at
$\lambda=0.05$, with the EA decreasing approximately twice as rapidly as the IP
(Tables~\ref{tab:ip}--\ref{tab:ea}).
The optical gap exhibits a smaller blueshift
($+0.033$~eV), leading to an increase of the finite-basis difference $E_b$ by $+0.011$~eV. 
Thus, within cc-pVDZ the quasiparticle gap opens more than the lowest optical gap.
The same trend is observed for HF ($+0.016$~eV) and CH$_4$
($+0.010$~eV), whereas NH$_3$ shows the opposite behavior ($-0.007$~eV).
The sign of the finite-basis $E_b$ shift
is thus molecule dependent and determined by the
competition between the opening of the fundamental quasiparticle gap $E_\mathrm{gap}$ and the blueshift of the optical excitation $\Omega_{S_1}$.
The singlet--triplet splitting also increases for all five molecules, by
$+5$ to $+19$~meV at $\lambda=0.05$, because the triplet shifts less than
the singlet. Such changes may be relevant to polariton-modified
intersystem-crossing and singlet--triplet
engineering~\cite{Stranius2018,Eizner2019}. A coupling-strength ($\lambda$) scan for H$_2$O
(Fig.~\ref{fig:bindcompare}, top left) confirms that all cavity shifts
follow the expected quadratic scaling with $\lambda$
up to $\lambda=0.1$, yielding $\delta E_\mathrm{gap}=+0.171$~eV
and $\delta E_b=+0.042$~eV.

\subsection{Cavity-kernel transparency and an empirical polarization criterion}\label{sec:exciton-selrule}

The channel decomposition provides a microscopic interpretation of the cavity-induced binding shifts of Table~\ref{tab:exciton}.
As discussed in Sec.~\ref{sec:bsetheory},
the cavity modifies the optical excitation through three identifiable ingredients:
(i) the cavity-dressed quasiparticle
energies entering the BSE diagonal (the reference channel, ``QP''), 
(ii) the DSE $d\otimes d$ contribution to the BSE kernel (``DSE''), 
and (iii) the photon-mediated contribution to 
the screened interaction (``photon''). 
Because BSE eigenvalues depend nonlinearly on the kernel, changes
obtained by independently toggling matrix terms are not additive in
general. We therefore define the total kernel residual unambiguously as
the full result minus the QP-only result. The separate DSE and photon
numbers below are diagnostic toggle calculations performed at the same
fixed cavity-dressed QP energies; for H$_2$O they cancel within
numerical precision:
\begin{align*}
\delta_\lambda\Omega_{S_1}(\lambda{=}0.05)
&=\underbrace{0.0332}_{\mathrm{QP}}\:
  \underbrace{-0.0105}_{\mathrm{DSE}}\:
  \underbrace{+0.0105}_{\mathrm{photon}}
  =0.0332~\mathrm{eV},\\[2pt]
\delta_\lambda\Omega_{S_1}(\lambda{=}0.10)
&=0.1294-0.0392+0.0392=0.1294~\mathrm{eV}.
\end{align*}
The two diagnostic changes are equal in magnitude and
opposite in sign, cancelling to $10^{-6}$~eV in this calculation.
Consequently, the optical-gap shift arises numerically from the
quasiparticle channel. The bare-photon identity of
Sec.~\ref{sec:bsetheory} explains the cancellation of the matching DSE
term inside $W$; it does not eliminate the separate exchange-channel
DSE term. Polariton dressing and the remaining exchange term can
therefore produce residual corrections in electron--hole channels with
the appropriate symmetry and dipole coupling.
For minimal-basis H$_2$O at $\lambda=0.1$, 
only 64 of the 1600 kernel elements are modified, shifting a handful of
high-lying excitations by as much as $0.3$~eV, while 
low-lying excitons remain unchanged ($<10^{-5}$~eV). 
Consequently, every observable reported in
Table~\ref{tab:exciton} for H$_2$O is inherited from the quasiparticle channel:
\emph{the screening-folded QED-BSE kernel is effectively cavity-transparent for the optical gap}.
This transparency is not universal. In the present
five-molecule data set, a useful empirical criterion is that an
appreciable residual reaches the lowest exciton only when it is both
bright along the cavity polarization and strongly $z$-dipolar; only
NH$_3$ meets both conditions. Because a general BSE root contains many
orbital transitions, this pattern should be regarded as a
data-supported polarization criterion rather than a general selection
rule derived solely from the bare-photon cancellation.

\begin{table}[!htbp]
\centering
\caption{Channel decomposition of the cavity-induced exciton-binding
shift $\delta_\lambda E_b$ (meV; full QED-BSE@QED-evGW, cc-pVDZ,
$\lambda=0.05$, $z$-polarized) into the quasiparticle channel
$\delta_\lambda E_b^\mathrm{elec}$ (cavity-dressed QP energies, both
kernel cavity channels off) and the residual of the DSE--photon
cancellation $\delta_\lambda E_b^\mathrm{ker}=\delta_\lambda E_b
-\delta_\lambda E_b^\mathrm{elec}$. $\mu_z^2(S_1)$ is the squared
$z$-component of the $S_1$ transition dipole and $\Delta d_z=d^z_{LL}
-d^z_{HH}$ the change in permanent $z$-dipole on the HOMO$\to$LUMO
excitation (a.u.).  Within the molecular set examined, the
kernel residual is appreciable only when $S_1$ is both $z$-bright
($\mu_z\neq0$) and strongly $z$-dipolar ($\Delta d_z$ large),
conditions satisfied only by NH$_3$.} \label{tab:excitonchannel}
\begin{ruledtabular}
\begin{tabular}{lccccc}
system & $\mu_z^2(S_1)$ & $\Delta d_z$ &
$\delta_\lambda E_b^\mathrm{elec}$ & $\delta_\lambda E_b^\mathrm{ker}$ &
$\delta_\lambda E_b$ \\
\hline
H$_2$O & $0.00$ & $-1.34$ & $+10.5$ & $\phantom{-}0.0$ & $+10.5$ \\
HF     & $0.00$ & $+2.03$ & $+16.3$ & $\phantom{-}0.0$ & $+16.3$ \\
NH$_3$ & $0.27$ & $-1.21$ & $+11.0$ & $-16.9$ & $-5.9$ \\
CH$_4$ & $0.25$ & $-0.37$ & $+10.3$ & $\phantom{-}0.0$ & $+10.3$ \\
PH$_3$ & $0.00$ & $-0.82$ & $+24.8$ & $\phantom{-}0.0$ & $+24.8$ \\
\end{tabular}
\end{ruledtabular}
\end{table}

\begin{figure*}[!htbp]
\centering
\includegraphics[width=0.8\textwidth]{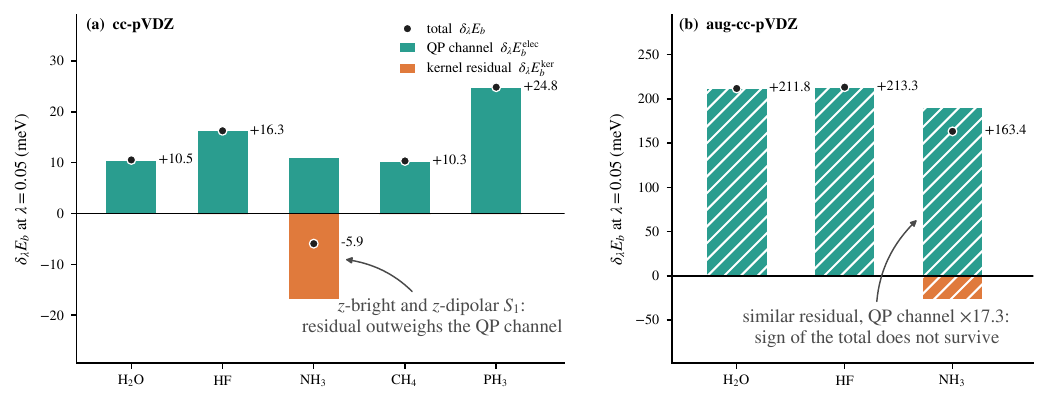}
\caption{Cavity-induced change of the $S_1$ exciton binding energy at
$\lambda=0.05$, resolved into the two channels of
Table~\ref{tab:excitonchannel} (full QED-BSE@QED-evGW, cc-pVDZ,
$z$-polarized). Bars: the quasiparticle channel
$\delta_\lambda E_b^\mathrm{elec}$ (both kernel cavity channels off) and the residual
$\delta_\lambda E_b^\mathrm{ker}$ of the DSE--photon cancellation; symbols: their sum.
(a)~cc-pVDZ: the residual is nonzero only for NH$_3$, the only molecule whose $S_1$ is
both $z$-bright and strongly $z$-dipolar, and there it outweighs the
positive quasiparticle channel. 
(b)~aug-cc-pVDZ (hatched): the residual is basis-robust ($-16.9\to-26.8$~meV) but the
quasiparticle channel grows by an order of magnitude ($11.0\to190.2$~meV, $\times17.3$), so the NH$_3$
sign reversal does not survive. 
The empirical polarization
criterion describes the residual, not the sign of the total
(Sec.~\ref{sec:exciton-basis}).
}
\label{fig:binding}
\end{figure*}

The polarization pattern helps rationalize the
molecule-dependent residual in Table~\ref{tab:exciton}; it does not by
itself determine the sign of $\delta_\lambda E_b$.
Extending the channel
decomposition for all five molecules
(Table~\ref{tab:excitonchannel} and Fig.~\ref{fig:binding}) separates
$\delta_\lambda E_b$ into the
quasiparticle contribution, $\delta_\lambda E_b^\mathrm{elec}$, 
obtained from cavity-dressed QP energies with both kernel channels disabled,
and the residual kernel contribution,
$\delta_\lambda E_b^\mathrm{ker}$ arising from the incomplete cancellation between the DSE--photon terms.
The quasiparticle contribution is positive for every molecule
($+10$ to $+25$~meV): the cavity enlarges the quasiparticle gap, thereby reducing the electronic polarizability and
screening and making the attractive direct interaction stronger in
this finite-basis calculation.
If the kernel were completely transparent, 
every molecule would exhibit this positive binding shift.  
The DSE and photon contributions largely cancel through the 
$\mp d\otimes d$ identity, leaving only a residual 
that acts within electron--hole channels coupled to the cavity polarization ($z$-dipole). 
In the present molecular set, this residual is appreciable
for the lowest exciton only when that state is bright along the cavity
axis $z$ and also has a large dipolar descriptor $\Delta d_z$.
For H$_2$O and HF, the lowest exciton is the out-of-plane  ($\pi$-type)
transition orthogonal to the cavity ($\mu_z^2(S_1)=0$, enforced by
$C_{2v}$/$C_{\infty v}$ symmetry), so the residual vanishes and
$\delta_\lambda E_b\simeq\delta_\lambda E_b^\mathrm{elec}>0$; CH$_4$
possesses a $z$-component within its degenerate $T_2$ exciton, but its
non-polar $T_d$  symmetry ($\Delta d_z=-0.4$~a.u.) leaves only a negligible residual
($\sim\!0$~meV), so the exciton again binds more strongly.
PH$_3$ provides the decisive counterexample showing 
that molecular polarity alone is insufficient. 
Although PH$_3$ is polar $C_{3v}$,
its lowest exciton is a $z$-\emph{dark} $E$-symmetry valence state
($\mu_z^2(S_1)=0$), preventing the residual from coupling to it
($\delta_\lambda E_b^\mathrm{ker}=0$).
Consequently, the positive quasiparticle contribution alone
($+25$~meV) determines the binding shift.
NH$_3$, by contrast, satisfies both parts of the empirical criterion.
Its lowest exciton is the nitrogen lone-pair transition polarized
\emph{along} the $C_3=z$ axis ($\mu_z^2(S_1)=0.27$),
and its frontier orbitals carry a large transition
$z$-dipole ($\Delta d_z=-1.2$~a.u.).
The polariton-dressing residual therefore acts directly on the
lowest exciton $S_1$. A first-order decomposition of the kernel attributes
the entire residual to
the single diagonal HOMO$\to$LUMO matrix element, corresponding to the lone-pair
exciton's own $z$-bright channel, while the remaining $\sim10^2$
matrix elements contribute less than $1$~meV in total.
This residual ($-16.9$~meV) overwhelms
the positive  quasiparticle contribution ($+11.0$~meV), so the exciton binding
\emph{weakens} ($\delta_\lambda E_b=-5.9$~meV).

{In this basis, the residual kernel correction is negligible for 
the lowest excitons of H$_2$O, HF, CH$_4$, and PH$_3$, 
each of which fails at least one part of the bright/dipolar criterion,
whereas it is negative for the lowest exciton of NH$_3$, 
which satisfies both parts.
This empirical pattern is robust across the two bases tested, but it is not a general
theorem following from the bare-photon identity alone. Whether the
residual overcomes the quasiparticle channel and reverses the net
finite-basis shift is a separate, strongly basis-sensitive question,
as discussed in Sec.~\ref{sec:exciton-basis}.
}
Genuine photon-mediated electron--hole binding, arising from the exchange of a cavity
photon between electron and hole at $\omega\approx\tilde\omega$, 
is precisely the contribution that is projected out when the photon is integrated into a static $W$.
It is recovered only when the photon is retained as an explicit
degree of freedom, as in the polaritonic QED-BSE formulation of
Sec.~\ref{sec:bsetheory}.

The most direct test of this transparency is to point the cavity at
the exciton itself. Retuning $\omega_\mathrm{cav}$ from the
RPA-resonant $11.31$~eV down to the BSE optical gap,
$\omega_\mathrm{cav}=\Omega_{S_1}(\lambda{=}0)=7.937$~eV, and
repeating the scan yields a quantified null result. Every shift in
Fig.~\ref{fig:bindcompare} (water) changes by at most $2\%$
($\delta_\lambda\Omega_{S_1}=+0.0335$ vs $+0.0332$~eV,
$\delta_\lambda E_b=+0.0104$ vs $+0.0105$~eV at $\lambda=0.05$; solid
versus dashed curves). The $\lambda$-dependence remains a smooth
$\lambda^2$ curve with no resonance feature whatsoever, and the kernel
decomposition cancels channel by channel exactly as before
($\mp0.0104$~eV at $\lambda=0.05$, $\mp0.0391$~eV at $\lambda=0.10$).
The screening-folded excitation space explains the null: it
contains no photon configuration, so an exciton-resonant cavity
has nothing to
anticross with. The static folding removes the photon identically at
\emph{any} detuning, and the residual tuning dependence enters only
through the quasiparticle energies, whose cavity shifts are dominated
by the $\omega_\mathrm{cav}$-independent mean-field DSE
(Sec.~\ref{sec:gw-bench}). Rabi splitting of the optical response,
the defining polaritonic phenomenon already evident in the QED-RPA
absorption spectrum of Fig.~\ref{fig:abs}, is absent from
the screening-folded BSE.
It is recovered only when the photon is restored
as an explicit degree of freedom in the excitation space, as in the
polaritonic QED-BSE of Sec.~\ref{sec:bsetheory} and Fig.~\ref{fig:abs}(b).
The screening-folded
QED-BSE describes how the cavity dresses the exciton's ingredients, not
how exciton and photon hybridize. Figure~\ref{fig:bindcompare} places the two contrasting
cases side by side, each with the cavity tuned to its own optical gap:
the channel decomposition is cavity-transparent for water (the DSE and
photon channels cancel, so $\delta_\lambda E_b$ follows the
quasiparticle channel and the exciton binds more), but not for NH$_3$,
whose $z$-bright $S_1$ leaves an uncancelled DSE residual that overturns
the quasiparticle binding and weakens the exciton binding.

{
\subsection{Basis-set robustness of the channel
decomposition}\label{sec:exciton-basis}

All values in Tables~\ref{tab:exciton} and~\ref{tab:excitonchannel} are
computed in cc-pVDZ, where the LUMO represents an artificially bound, diffuse attachment state whose properties are strongly basis-dependent. This matters here more than it does
for the charged excitations, because the two channels of the
decomposition respond to the basis in completely different ways. 
Since $\delta_\lambda E_\mathrm{gap}=\delta_\lambda\mathrm{IP}
-\delta_\lambda\mathrm{EA}$ and $\delta_\lambda\mathrm{EA}$ grows by a factor of three upon adding diffuse functions
(Sec.~\ref{sec:gw-bench}: $-0.075\to-0.235$~eV for water at
$\Delta$QED-CCSD), while $\delta_\lambda\mathrm{IP}$ remains basis-stable at $-0.02$~eV, 
the fundamental-gap shift, and consequently the quasiparticle
channel $\delta_\lambda E_b^\mathrm{elec}$, must increase by roughly the same factor. 
Whether the kernel residual $\delta_\lambda E_b^\mathrm{ker}$, a matrix-element effect, grows with
it is not something the cc-pVDZ data can answer. We therefore repeated
the full construction in aug-cc-pVDZ
(Table~\ref{tab:excitonbasis}) for the three molecules that decide the
question: the two whose $S_1$ is $z$-dark by symmetry, H$_2$O and HF,
which fix the transparent baseline, and NH$_3$, the one
transparency-breaking case. CH$_4$ and PH$_3$ were not repeated in the
diffuse basis; they add a third and fourth instance of the transparent
baseline rather than new information.

\begin{table*}[tb]
\centering
\caption{Basis-set dependence of the
neutral-excitation observables and of the channel decomposition (full
QED-BSE@QED-evGW, $\omega_\mathrm{cav}=0.415668$~$E_h$,
$z$-polarized, $\lambda=0.05$). $\delta_\lambda X=X(0.05)-X(0)$;
energies in eV, channels in meV, $\Delta d_z$ in a.u. The kernel
residual is nonzero only for NH$_3$ in \emph{both} bases, 
showing that the observed bright/dipolar criterion is stable across these two bases,
whereas the quasiparticle channel,
governed by $\delta_\lambda E_\mathrm{gap}$ and hence by the
basis-sensitive attachment channel, grows by an order of magnitude upon inclusion of diffuse functions. 
The cc-pVDZ rows reproduce Tables~\ref{tab:exciton} and~\ref{tab:excitonchannel} entry for entry;
the aug-cc-pVDZ NH$_3$ entry predates the frontier-gap convention of
Sec.~\ref{sec:abs}, which moves its cc-pVDZ counterpart by
$1.1$~meV.}\label{tab:excitonbasis}
\begin{ruledtabular}
\begin{tabular}{llcccccccccc}
system & basis & $E_\mathrm{gap}$ & $\delta_\lambda E_\mathrm{gap}$ &
$\Omega_{S_1}$ & $\delta_\lambda\Omega_{S_1}$ & $E_b$ &
$\delta_\lambda E_b$ & $\delta_\lambda E_b^\mathrm{elec}$ &
$\delta_\lambda E_b^\mathrm{ker}$ & $\mu_z^2(S_1)$ & $\Delta d_z$ \\
\hline
H$_2$O & cc-pVDZ & 16.383 & $+0.044$ & 7.937 & $+0.033$ & 8.447 & $+0.011$ & $+10.5$ & $\phantom{-}0.0$ & 0.00 & $-1.34$ \\
 & aug-cc-pVDZ & 13.000 & $+0.200$ & 7.306 & $-0.012$ & 5.694 & $+0.212$ & $+211.8$ & $\phantom{-}0.0$ & 0.00 & $-1.65$ \\
\hline
HF & cc-pVDZ & 20.191 & $+0.034$ & 10.837 & $+0.018$ & 9.354 & $+0.016$ & $+16.3$ & $\phantom{-}0.0$ & 0.00 & $+2.03$ \\
 & aug-cc-pVDZ & 16.555 & $+0.219$ & 10.280 & $+0.006$ & 6.275 & $+0.213$ & $+213.3$ & $\phantom{-}0.0$ & 0.00 & $+2.32$ \\
\hline
NH$_3$ & cc-pVDZ & 15.185 & $+0.053$ & 7.824 & $+0.059$ & 7.361 & $-0.006$ & $+11.0$ & $-16.9$ & 0.27 & $-1.21$ \\
 & aug-cc-pVDZ & 11.595 & $+0.193$ & 6.676 & $+0.029$ & 4.919 & $+0.163$ & $+190.2$ & $-26.8$ & 0.46 & $-1.44$ \\
\end{tabular}
\end{ruledtabular}
\end{table*}

Three features are basis-robust, whereas one is not. 
\emph{(i)} The empirical polarization criterion: $\delta_\lambda E_b^\mathrm{ker}$ vanishes for H$_2$O and
HF in \emph{both} bases, as required by the $z$-darkness of their $S_1$ states,
and is nonzero only for NH$_3$, consistent with the
$z$-bright/$z$-dipolar descriptor. 
In cc-pVDZ, the same behavior holds
for PH$_3$, and for CH$_4$ within the $0.5$~meV ambiguity associated with
assigning $S_1$ within its $z$-split $T_2$ manifold.
This behavior follows from the symmetry of
$S_1$ rather than the basis size.
This supports the
robustness of the observed residual pattern, while the exact identity
of Appendix~\ref{ap:cancel} remains limited to the matching DSE and
bare-photon terms inside $W$.
\emph{(ii)} The sign and approximate magnitude of the residual are also robust where it survives: for NH$_3$,
the residual increases from $-16.9$ to $-26.8$~meV, tracking the increase in the two
quantities identified 
empirically as useful descriptors,
$\mu_z^2(S_1)$ from
$0.27$ to $0.46$ and $\Delta d_z$ from $-1.21$ to $-1.44$~a.u.
\emph{(iii)} Finally, the quasiparticle channel remains  
positive  for every molecule in both bases.

What is \emph{not} robust is the magnitude of that quasiparticle
channel, and therefore the sign of the total. In aug-cc-pVDZ
$\delta_\lambda E_b^\mathrm{elec}$ grows by a factor of $13$--$20$
($+10.5\to+211.8$~meV for water, $+11.0\to+190.2$~meV for NH$_3$),
tracking the growth of $\delta_\lambda E_\mathrm{gap}$
($+0.044\to+0.200$~eV for water) as anticipated, while the optical-gap
shift $\delta_\lambda\Omega_{S_1}$ does \emph{not} follow it at all: it
roughly halves for NH$_3$ ($+0.059\to+0.029$~eV) and for water it
\emph{changes sign}, from a $+0.033$~eV blueshift in cc-pVDZ to a
$-0.012$~eV redshift in aug-cc-pVDZ. The
$-26.8$~meV NH$_3$ residual is therefore no longer competitive with a
$+190$~meV quasiparticle channel, and the sign reversal seen in
cc-pVDZ does not survive: in aug-cc-pVDZ the cavity strengthens the
electron--hole interaction for NH$_3$ as well
($\delta_\lambda E_b=+163$~meV).

The consequence for the central claim is as follows. 
The bright/dipolar criterion consistently identifies the molecules with a
non-negligible kernel residual in the two bases tested, but the present
data do not establish a universal selection rule or determine the sign
of the total shift. That sign is decided by the competition between
the residual and the quasiparticle channel.
The quasiparticle channel is dominated by an electron-attachment channel, which is not
converged in either basis considered here because the anion is unbound in both cases. 
Accordingly, the sign reversal observed for NH$_3$ should be interpreted as evidence
that the kernel residual can be \emph{comparable} in magnitude to the quasiparticle channel.
The physically substantive result is that the
residual, which measures the cavity dependence remaining in the two-particle kernel after the QP-only contribution is removed, can be comparable in magnitude to the quasiparticle channel. 
It should not be interpreted as a converged prediction that a cavity unbinds the lowest excitation of ammonia. 
Establishing such a prediction requires a converged treatment of the electron-attachment channel, 
for example through a genuinely
bound anion, a continuum-corrected basis, or the periodic setting in
which Eq.~\eqref{eq:Eb} recovers its usual meaning as an exciton
binding energy.
}

\section{Conclusions}\label{sec:conclusion}

We have organized the quantum-electrodynamical contributions to the
electron self-energy of a cavity-coupled molecule into three channels:
the static DSE exchange residing in the QED-HF reference, the direct
DSE augmentation of the screened interaction, and the polariton-pole
channel that carries the bilinear electron--photon coupling into $\Sigma_c$.  
The resulting quasiparticle IPs and EAs of four ten-electron hydrides and
two aromatic hydrocarbons (benzene and naphthalene) were benchmarked
against the cavity $\Delta$-method. There the vertical IP and EA are
ground-state energy differences of the same Pauli--Fierz Hamiltonian
evaluated across the three charge sectors.
The $\Delta$-method ladder ($\Delta$QED-HF $\to$
$\Delta$QED-CCSD $\to$ $\Delta$QED-DMRG $\to$ exact $\Delta$QED-FCI in
a minimal basis) is internally consistent among its correlated rungs
for both observables.
The $\Delta$QED-CCSD and $\Delta$QED-FCI results differ by  at most $2$~meV in the minimal basis, 
while the $\Delta$QED-CCSD and the near-exact $\Delta$QED-DMRG cavity-induced
shifts in cc-pVDZ agree within $1$~meV. 
This agreement provides a near-exact numerical yardstick
for the cavity-induced shifts in the stated electronic bases and
photon truncations; the individual CCSD and finite-bond-dimension DMRG
calculations remain approximations.

Our central numerical finding is a clean separation of scales.
The \emph{absolute} charged excitations of a cavity-coupled molecule are described by QED-$GW$ about as accurately as ordinary $GW$@HF describes the \emph{bare} molecule: evGW agrees with $\Delta$QED-CCSD within $0.0$--$0.25$~eV for the IPs and within $0.05$--$0.15$~eV for the EAs.
By contrast, the \emph{cavity-induced shifts} of the IPs --- the observable specific to polaritonic chemistry --- are generally overestimated in magnitude. 
The overestimation ranges from $14\%$\ to $57\%$ across the hydrides and reaches $42\%$ for resonant water, $69\%$ for benzene, and $95\%$ for water in aug-cc-pVDZ. 
One-shot G$_0$W$_0$ is an exception for HF, 
recovering only about half the magnitude of the correlated shift. 
In this case, the shift results from the difference between two small, competing contributions, whose balance is captured only after eigenvalue self-consistency.
The attachment shifts are reproduced to a
few to ten per cent. Benzene sharpens the diagnosis: its absolute $\pi$ IP agrees with
$\Delta$QED-CCSD to within $10$~meV, yet its cavity-induced shift retains the
same $\sim$60--70\% overestimate.
The channel comparison shows that a correlated cavity
contribution partially offsets the mean-field DSE penalty and that the
ring-level $\Sigma_c$ recovers only part of it. The pattern is
consistent with the size of the difference in dipole fluctuations
between the two charge states: for the closed-shell molecules studied
here the larger error occurs in ionization, whereas for the ionic
sodium halides with bound, polarizable anions it occurs in attachment
[Appendix~\ref{ap:sodium_halides}]. This is evidence for missing
vertex or differential-correlation physics beyond ring screening, not
a unique diagrammatic identification of an electron--photon counter-term.

Two corollaries complete the picture. 
First, the polariton-frequency physics that static
approximations discard is directly observable. The frequency-resolved
spectral function of water develops a photoemission sideband at
$\varepsilon_\mathrm{HOMO}^\mathrm{QED\text{-}HF}-\tilde\omega$, in a spectral window that is
numerically empty at this G$_0$W$_0$ level for the bare molecule, with pole strength scaling as
$\lambda^2$. This produces an isolated, falsifiable signature of the
bilinear electron--photon coupling within the present approximation.
Second, a QED Bethe--Salpeter equation built on the QED-$GW$ quasiparticles delivers gaps and a finite exciton binding energy within one framework. 
The cavity-induced change in the finite-basis difference
$E_b=E_\mathrm{gap}-\Omega_{S_1}$ is partitioned into a QP-only result
and a kernel residual defined by subtraction. At $\omega=0$, exchange
of a bare photon cancels the matching DSE term inside the direct
screened interaction $W$ (Appendix~\ref{ap:cancel}); it does not cancel
the separate exchange-channel DSE term in the complete BSE kernel. In
the present data set, the residual is negligible for H$_2$O and HF,
whose lowest states are $z$-dark, for CH$_4$, which is $z$-bright but
has a small dipolar descriptor, and for PH$_3$, whose lowest state is
$z$-dark. NH$_3$ is the only examined case with both a $z$-bright state
and a large $\Delta d_z$, and its negative residual changes from
$-16.9$ to $-26.8$~meV between the two bases. This is an empirical,
basis-stable criterion within the molecular set, not a universal
selection rule derived from the static identity alone.

The quasiparticle channel is positive throughout but is controlled by
$\delta_\lambda E_\mathrm{gap}$ and therefore by the poorly converged
attachment channel of these unbound anions. It grows from
$+10$--$25$~meV in cc-pVDZ to $+190$--$213$~meV in aug-cc-pVDZ. The
kernel residual can therefore be comparable to the QP contribution in
cc-pVDZ but not in the diffuse basis. These quantities should be read
as finite-basis electron--hole interaction diagnostics, not as
converged molecular exciton-binding shifts.
The screening-folded kernel produces no anticrossing, retuning the cavity into
exact resonance changing the shifts by $\le2\%$, the photon being absent from
the folded excitation space; the Rabi splitting is recovered by keeping the
photon explicit, as in the polaritonic QED-BSE of
Sec.~\ref{sec:bsetheory}.

Independent analyses of charged and neutral excitations yield a consistent picture:
the characteristic physics of a molecule coupled to a resonant cavity is governed
by processes occurring near the cavity frequency, $\omega\approx\omega_\mathrm{cav}$.
The polariton photoemission sideband and the explicit-photon polaritonic
QED-BSE both place this physics in observables, and the $\Delta$-method
benchmark quantifies where the ring-level self-energy still falls short.
Frequency-resolved many-body treatments, vertex corrections beyond the
ring-level screening~\cite{Cunningham2018}, and the strong-coupling
(non-perturbative) regime remain the natural next steps toward a complete
description of cavity-modified molecular systems.

\begin{acknowledgments}
CWM acknowledges the support provided by the National Research Foundation of Korea (NRF) grants funded by the Korean government (MSIT) (Grant No.~RS-2022-NR072058, RS-2026-25501632). 
SYW and CWM acknowledge the support provided by the NRF grants funded by the Korean government (MSIT) (Grant No.~RS-2024-00407680). 
THP, GBS, and CWM acknowledge the support provided by the Korea Institute of Energy Technology Evaluation and Planning (KETEP) and the Ministry of Climate, Energy and Environment (MCEE) of the Republic of Korea (No. RS-2025-25459387).
DCY acknowledges the support provided by the NRF grant RS-2026-25576504.
SYW acknowledges the support provided by the NRF grant RS-2026-25577169.
TIK acknowledges the support provided by the Korea Institute of Energy Technology Evaluation and Planning(KETEP) and the Ministry of Climate, Energy \& Environment(MCEE) of the Republic of Korea (No. RS-2026-25534494).
The authors are grateful for the computational resources at the Korea Institute of Science and Technology Information~(KISTI) with the \textsc{Nurion} cluster (KSC-2025-CRE-0286, KSC-2025-CRE-0316, KSC-2025-CRE-0125, KSC-2025-CRE-0122). Computational work for this research was also partially performed on the \textsc{Olaf} cluster supported by IBS Research Solution Center and on the GPU cluster supported by NIPA.
MM, JB, and LV would like to acknowledge support by the Ministry of Education of the Czech Republic MSMT-CR, INTER-EXCELLENCE II, INTER-ACTION [LUAUS25286], through the e-INFRA CZ (ID:90254), and the Advanced Multiscale Materials for Key Enabling Technologies project, Project No. CZ.02.01.01/00/22\_008/0004558, Co-funded by the European Union.
\end{acknowledgments}

\section*{Code and data availability}

To support transparency and reproducibility, the implementation, example scripts, 
and associated reproducibility artifacts have been made publicly available 
through the OmegaQMC repository at
\url{https://github.com/myung-group/OmegaQMC}. 
The repository include a \texttt{examples/qed\_gw/README.md} file 
describing the available example scripts, their usage, and the procedures required to reproduce the principal results reported in this work, including the benchmark calculations, exciton-binding-energy analysis, spectral-function calculations, and polaritonic QED-BSE--FCI validation.
The MOLMPS program used for the QED-DMRG calculations is publicly available at \url{https://gitlab.com/molmps/scalable}. 

\appendix

\begin{table*}[tb]
\centering
\caption{Cavity-modulated IPs and EAs of the sodium halides using the setup
of Ref.~\onlinecite{DePrince2021} (def2-TZVPPD at the B3LYP geometries
$r_e(\mathrm{NaF})=1.936$ and $r_e(\mathrm{NaCl})=2.369$~\AA,
$\omega_\mathrm{cav}=2.0$~eV along the molecular axis). Upper rows of
each block: absolute values at $\lambda=0$ (eV); lower rows:
cavity-induced shifts $\delta_\lambda X=X(0.05)-X(0)$ (meV). Columns
labelled ``Ref.'' are the published values, tabulated to two decimals
and so carrying a $\pm10$~meV rounding uncertainty on the shifts.}\label{tab:nax}
\begin{ruledtabular}
\begin{tabular}{llccccc}
 & & \multicolumn{2}{c}{$\Delta$QED-HF} & \multicolumn{2}{c}{QED-$GW$ (this work)} & $\Delta$QED-CC \\
\cline{3-4}\cline{5-6}
system & quantity & this work & Ref. & G$_0$W$_0$ & evGW & Ref. \\
\hline
NaF & IP & 8.10 & 8.10 & 9.97 & 9.70 & 9.96 \\
 & EA & 0.37 & 0.37 & 0.41 & 0.41 & 0.43 \\
 & $\delta_\lambda$IP & $-44$ & $-50$ & $+28$ & $-39$ & $-40$ \\
 & $\delta_\lambda$EA & $-241$ & $-240$ & $-391$ & $-286$ & $-170$ \\
\hline
NaCl & IP & 7.99 & 7.99 & 9.16 & 9.13 & 9.00 \\
 & EA & 0.56 & 0.56 & 0.61 & 0.61 & 0.64 \\
 & $\delta_\lambda$IP & $-86$ & $-90$ & $-22$ & $-80$ & $-60$ \\
 & $\delta_\lambda$EA & $-227$ & $-220$ & $-502$ & $-350$ & $-150$ \\
\end{tabular}
\end{ruledtabular}
\end{table*}

\section{Physical regime: single-molecule versus collective
coupling}\label{ap:regime}

It is useful to state explicitly what $\lambda=0.05$~a.u.\ implies,
since this value determines the physical regime in which all
results in the main text should be interpreted. Using the standard relation
$\lambda=\sqrt{1/(\epsilon_0 V)}$ between the coupling magnitude
and an effective mode volume $V$~\cite{Haugland2020}, 
$\lambda=0.05$ corresponds to 
$V=4\pi/\lambda^2\simeq5.0\times10^3,a_0^3=0.74$~nm$^3$, 
or a cubic mode volume with a side length of approximately $0.9$~nm. 
For $\lambda=0.10$, the corresponding volume is $0.19$~nm$^3$.
These mode volumes are characteristic of picocavities 
and plasmonic nanogaps rather than conventional Fabry--P\'erot cavities.
For comparison, a planar cavity resonant at $\omega_\mathrm{cav}=11.31$~eV 
has a free-space wavelength of approximately $110$~nm and a diffraction-limited mode volume of order
$(\lambda_\mathrm{ph}/2)^3\approx1.6\times10^5$~nm$^3$. 
The corresponding \emph{single-molecule} coupling is only 
$\lambda\sim10^{-4}$~a.u. Thus, the results presented here should
be interpreted as single-molecule strong-coupling results,
relevant to plasmonic picocavities in which single-molecule strong
coupling has been experimentally demonstrated~\cite{Chikkaraddy2016}. 
The ultrastrong-coupling label, when applicable to a specific transition, requires a separate comparison of $g$ with $\omega_\mathrm{cav}$.
They should not be interpreted as a simulation of a molecular
ensemble in a conventional Fabry--P\'erot cavity, which is the
regime addressed by much of the experimental literature and 
in which the observed vacuum Rabi splitting is primarily associated
with the collective $\sqrt{N}$ enhancement of the bright state. 
Microscopic derivations based on macroscopic QED further show that direct
intermolecular interactions must be retained when reducing such
many-molecule systems to effective exciton--polariton models~\cite{ChuangHsu2024Microscopic}.
At the diffraction-limited planar-cavity mode volume just estimated, approximately
$N\sim2\times10^5$ molecules would be required to obtain 
a collective coupling of $0.05$~a.u.

We note that the structure of Eq.~\eqref{eq:channels} already
indicates an important distinction in their scaling with $N$. 
The bilinear light--matter vertex $\bm g$, and consequently 
the polariton-pole channel $M^\mathrm{ph}$, is a collective
quantity: it couples the photon to the bright molecular
superposition and is enhanced by $\sqrt{N}$. 
In contrast, the DSE channel $M^\mathrm{DSE}$ and the implicit
reference channel are one-molecule operators. 
Their intermolecular cross terms are expected to cancel against 
the corresponding intermolecular Coulomb contribution~\cite{Fiechter2026}. 
Consequently, the three channels do not obey the same $N$-scaling,
and a per-molecule observable, such as the ionization potential of
an individual molecule in an ensemble, need not inherit the 
$\lambda^2$-scale magnitudes reported here.

What should transfer more directly to the collective regime is therefore not the numerical magnitude of the individual channel contributions, but the \emph{diagnostic structure} of the decomposition: namely, identifying which channel generates a given cavity-induced shift and determining where a ring-level self-energy fails to reproduce that contribution.

\section{$\Delta$QED-CCSD for NaF and NaCl}\label{ap:sodium_halides}

DePrince~\cite{DePrince2021} studies sodium halides rather than
hydrides, so we reproduce its setup exactly: NaF and NaCl in
def2-TZVPPD at the B3LYP/def2-TZVPPD equilibrium geometry, 
with $\omega_\mathrm{cav}=2.0$~eV polarized along the molecular axis, and
$\lambda$ varied from $0$ to $0.05$.
We augment the published results with QED-$GW$ data in Table~\ref{tab:nax}. 
The $\Delta$QED-HF column provides a stringent implementation check, since it is a mean-field quantity fixed by the
Pauli--Fierz Hamiltonian, coherent-state reference, and gauge origin. 
Our values agree with the published results to within $5$~meV  for both molecules and both observables at every coupling strength,
confirming that the QED-$GW$ results use the same convention.

The sodium halides also provide a useful contrast to the hydrides.
Because NaF and NaCl are strongly ionic (e.g., Na$^+$F$^-$) and  support \emph{bound} anions (e.g., NaF$^-$),
the change in dipole fluctuations upon electron attachment  is larger than that upon electron detachment.
Consequently, the neutral/anion pair exhibits the larger dipole-fluctuation difference than the neural/cation pair, reversing the ordering of the $GW$ errors.  
For NaF, evGW reproduces IP shift, $-39$~meV versus $-40$~meV from the $\Delta$QED-CC reference,
but substantially overestimates the magnitude of the EA shift, $-286$~meV versus $-170$~meV.
NaCl shows the same trend, with errors of $+34\%$ for the IP shift
($-80$ versus $-60$~meV) and $+134\%$ for the EA shift 
($-350$ versus $-150$~meV). 
Thus, as in the hydrides, the largest error occurs 
in the charge channel with the larger change in polarizability:
ionization for the hydrides, 
but electron attachment for these ionic species with bound, polarizable anions. 
This supports the interpretation that ring-level $\Sigma_c$
misses part of a correlated cavity contribution, while not uniquely identifying its diagrammatic origin.

The one-shot G$_0$W$_0$ result further illustrates the importance of eigenvalue self-consistency: for NaF it predicts a $+28$~meV blue shift of the IP, whereas the correlated references give a redshift. Eigenvalue self-consistency restores the balance between the Koopmans and correlation contributions. Finally, the $\Delta$QED-CC results are consistent with the EOM-EA-QED-CCSD-1 calculations of Ref.~\onlinecite{Liebenthal2022}, whose electron affinities agree with the $\Delta$QED-CCSD-1 values to within $0.01$~eV across the coupling strengths considered.

\section{Static cancellation of the photon exchange against the DSE kernel}\label{ap:cancel}

The index-specific identity used in Secs.~\ref{sec:abs} and~\ref{sec:exciton-selrule} is  that, at $\omega=0$, the exchange of a \emph{bare} cavity photon contributes $-d_{pq} d_{rs}$ and thereby cancels the matching
DSE term $+d_{pq} d_{rs}$ inside the direct interaction $W_{pq,rs}^\mathrm{stat}$.
It follows directly from the spectral representation of
Eq.~\eqref{eq:Wspec}.

To see this, switch off the electronic screening so that the photon mode of the
QED-RPA problem, Eq.~\eqref{eq:QEDRPA}, decouples from the
particle--hole block. The resulting $2\times2$ sub-problem has
$\mathcal{A}=\omega_\mathrm{cav}$, $\mathcal{B}=0$. 
Its positive-energy solution is $\Omega_\mathrm{ph}=\omega_\mathrm{cav}$ with 
$M=1$ and $N=0$, as fixed by the symplectic normalization of
Eq.~\eqref{eq:norm}; hence\ $(M+N)_\mathrm{ph}=1$. 
The corresponding transition density, Eq.~\eqref{eq:channels}, is purely photonic:
\begin{equation}
M_{pq,\mathrm{ph}}
=-\sqrt{\frac{\omega_\mathrm{cav}}{2}}\;d_{pq}\,(M+N)_\mathrm{ph}
=-\sqrt{\frac{\omega_\mathrm{cav}}{2}}\;d_{pq},
\label{eq:Mph_bare}
\end{equation}
which is the light--matter vertex $g_{pq}$ of Eq.~\eqref{eq:Abig}.
Substituting Eq.~\eqref{eq:Mph_bare} into the static screened
interaction of Eq.~\eqref{eq:Wstat}, the photon contribution is
\begin{align}
\Delta W^\mathrm{ph}_{pq,rs}
&=-\,2\,\frac{M_{pq,\mathrm{ph}}M_{rs,\mathrm{ph}}}{\Omega_\mathrm{ph}}
\;=\;-\,\frac{2}{\omega_\mathrm{cav}}\,
\frac{\omega_\mathrm{cav}}{2}\,d_{pq}d_{rs}\nonumber\\
&=\;-\,d_{pq}d_{rs}.
\label{eq:cancel}
\end{align}
Thus, the result is independent of $\omega_\mathrm{cav}$. 
Because $\bar v_{pq,rs}=(pq|rs)+d_{pq}d_{rs}$ [Eq.~\eqref{eq:Wspec}], 
the photon-mediated contribution cancels the DSE term identically, giving
\begin{equation}
    W^\mathrm{stat}_{pq,rs}=(pq|rs).
\end{equation}
Thus, in the absence of electronic screening, the direct
static interaction $W^\mathrm{stat}_{pq,rs}$ reduces to its Coulomb
part regardless of detuning. This statement concerns $W$ at the same
indices; the complete BSE kernel still contains the separate bare
exchange contribution $d_{ai}d_{bj}$ in
Eqs.~\eqref{eq:bse}--\eqref{eq:Bbse}.

The same result follows immediately in propagator language.
The photon-mediated electron--electron interaction is
$g_{pq}g_{rs}D(\omega)$, where the free-boson propagator is 
\begin{equation*}
    D(\omega)=\frac{2\omega_\mathrm{cav}}{(\omega^2-\omega_\mathrm{cav}^2)},
\end{equation*}
so that
$D(0)=-2/\omega_\mathrm{cav}$ 
and hence $g_{pq}g_{rs}D(0)=-d_{pq}d_{rs}$, reproducing Eq.~\eqref{eq:cancel}. 
The cancellation within the direct interaction is therefore
exact at any detuning for a bare photon. With electronic screening
restored, $\omega_\mathrm{cav}$ also enters through the true polariton
poles, their amplitudes, and interference among the electronic, DSE,
and photon transition-density components. Together with the separate
exchange-channel DSE term, these effects generate the residual kernel
changes examined in Sec.~\ref{sec:exciton-selrule}. Their projection
onto a BSE eigenstate depends on the full transition vector and its
symmetry; the bright/dipolar pattern observed here is therefore an
empirical criterion rather than a consequence of Eq.~\eqref{eq:cancel}
alone.

\bibliography{refs}

\end{document}